\documentclass[reqno]{amsart}

\theoremstyle{definition}

\numberwithin{equation}{section}
\usepackage{amsmath,amsfonts,amssymb,amsthm}
\usepackage{float}
\usepackage{enumitem}
\usepackage{comment}
\usepackage{overpic}
\usepackage{rotating}
\usepackage{xcolor}

\newcommand{\N}{\mathbb{N}}
\newcommand{\R}{\mathbb{R}}

\newcommand{\Z}{\mathbb{Z}}

\newcommand{\M}{\mathcal{M}}

\newcommand{\ep}{\varepsilon}

\DeclareMathOperator{\sech}{sech}

\makeatletter
\newcommand{\setword}[2]{%
  \phantomsection
  #1\def\@currentlabel{\unexpanded{#1}}\label{#2}%
}
\makeatother

\begin{document}

\title[Nonautonomous modelling in Energy Balance Models of climate.]{Nonautonomous modelling in Energy Balance Models of climate. Limitations of averaging and climate sensitivity}

\author[I.P.~Longo]{Iacopo P.~Longo}
\email[Iacopo P.~Longo]{iacopo.longo@imperial.ac.uk}
\author[R.~Obaya]{Rafael Obaya}
\email[Rafael Obaya]{rafael.obaya@uva.es}
\author[A.M.~Sanz]{Ana M.~Sanz}
\email[Ana M.~Sanz]{anamaria.sanz@uva.es}

\address[Iacopo P.~Longo]{Imperial College London,
Department of Mathematics, 180 Queen’s Gate
South Kensington Campus, SW7 2AZ London, United Kingdom.}
\address[Rafael Obaya]{Departamento de Matem\'{a}tica Aplicada,
Universidad de Valladolid,  Paseo Prado de la Magdalena 3-5, 47011 Valladolid, Spain.}
\address[Ana M.~Sanz]{Departamento de Didáctica de las Ciencias Experimentales, Sociales y de la Matemática, Universidad de Valladolid, Paseo de Belén 1, 47011 Valladolid, Spain.}
\address[Rafael Obaya \& Ana M.~Sanz]{IMUVA, Instituto de Investigaci\'{o}n en  Mate\-m\'{a}\-ti\-cas, Universidad de
Valladolid.}



\begin{abstract}
Starting from a classical Budyko-Sellers-Ghil energy balance model for the average surface temperature of the Earth, a nonautonomous version is designed by allowing the solar irradiance and the cloud cover coefficients to vary with time on a fast timescale, and to exhibit chaos in a precise sense. The dynamics of this model is described in terms of three existing nonautonomous equilibria, the upper one being attracting and representing the present temperature profile. The theory of averaging is used to compare the nonautonomous model and its time-averaged version. We analyse the influence of the qualitative properties of the time-dependent coefficients and obtain reasonable approximations close to the upper hyperbolic solution. Furthermore, previous concepts of two-point response and sensitivity functions are adapted to the nonautonomous context and used to value the increase in temperature when a forcing  caused by CO$_2$ and other emissions intervenes. 
\end{abstract}
\maketitle

\section{Introduction}
Energy balance models (EBMs) are conceptual in nature and provide a basic relation that describes the variation of the average surface temperature of the Earth in terms of a reduced number of climatic components with a relevant role. The interest and usefulness of these models lie in two complementary features; firstly, their simplicity allows a detailed analysis that provides essential qualitative features of the climate behaviour even over long periods of time, and secondly, the conclusions obtained using these models have been confirmed by studies carried out with much more sophisticated models, such as general circulation models (GCMs) (see  North et al.~\cite{north1981energy} and Ghil and Childress \cite{ghilchildress1987}, and more recent works such as Ghil and Lucarini \cite{ghil2020physics}, Geoffroy et al.~\cite{geoffroy2013transient} and Chao and Dessler \cite{Chao2021assessment}).

In contrast to physical evidence, EBMs are often studied as time-independent (i.e.~autonomous) systems  without clarifying the consequences of neglecting the external forcing in the modelling approach. Such choice is not always innocuous since nonautonomous differential equations can feature exceedingly complex dynamics (including chaos \cite{campos2023uniform,duenas2025Saddle}) already in dimension one. In this work we highlight the qualitative dynamical differences between an autonomous and a nonautonomous EBM and provide a rigorous framework within which an autonomous approximation is possible---albeit at the cost of losing significant dynamical information. 

We focus on a time-dependent version of the zero-dimensional energy balance model of  Budyko-Sellers-Ghil type given in \cite{budyko1969,sellers1969global,ghil1976climate, crafoord1978, gal1976energy}.
Its physical foundation is an energy conservation law between incoming solar radiation and outgoing emission that provides the instantaneous variation rate of the planet's average temperature times the constant of thermal inertia. 
The incoming solar radiation takes the form $\frac{1}{4} I(1-\alpha(T))$, where $I$ is the total solar irradiance and $\alpha(T) \in [0,1]$ is the average global albedo, a measure of the short-wave radiation reflected back into  space. 
The outgoing radiation is modelled by means of the Stefan-Boltzmann law, as the product of a multiple of the power 4 of the temperature times the effective emissivity, which represents the greenhouse effect and is parametrized as a function with range in $[0,1]$, decreasing with respect to the temperature and  depending on a cloud cover parameter $m$ that indicates the percentage of sky obscured by clouds on average.

The mentioned relevant climate components vary on a slow timescale. However, in most models in the literature (see~\cite{north1981energy}-\cite{Chao2021assessment}, \cite{budyko1969,sellers1969global,ghil1976climate})  they are assumed to be time independent, that is, the model is to be seen as a layer system of a more general model with slow time dependence. Moreover, physical considerations and experimental observations also justify their dependence on a fast timescale. During the period of a solar cycle, the levels of solar radiation and ejection of solar material exhibit a synchronous fluctuation from a period of minimum activity to a period of maximum activity, and then to a period of minimal activity. It is natural to assume a quasi-periodic expression to represent the solar irradiance $I(t)$ (see Kopp and Lean \cite{kopp2011new}). Moreover, the yearly global average cloud cover varies by approximately $3\%$ from year to year (see Stubenrauch et al.~\cite{stubenrauch2013assessment}). Among others, cycles lasting 15 days and others lasting six months have been detected in the behaviour of this variation (see~\cite{stubenrauch2013assessment}). In this paper, we model $m(t)$ by taking two different expressions. In the first one $m(t)$ is a quasi-periodic motion, while the second one takes into account the unpredictability of this coefficient by including a generic trajectory of a chaotic system. These are some basic arguments that justify the design of the EBM depending on the fast time as presented  in Section \ref{subsec:nonautonomous-forcing}.

Nonautonomous differential equations are investigated using the language of processes or dynamical methods based on the skew-product formalism. We adopt the latter approach in Section \ref{subsec:attr-repel-pair}, which requires a collective family of differential equations, endowed with an appropriate continuous flow. In this formulation, it is possible to transfer information between these equations using methods of ergodic theory and topological dynamics. The arguments of Longo et al.~\cite{longo2021rate,longo2024critical} permit us to prove that the proposed EBMs admit three nonautonomous equilibria that determine the global dynamics of the model. These equilibria are continuous functions, with dependence on time, that are capable of reproducing essential dynamic properties of those shown by the model coefficients, frequently given by a superposition of quasi-periodic effects with others that incorporate ingredients of complexity typical of chaotic dynamics.

At least intuitively, it seems reasonable to eliminate the time-dependent variation of the EBM on the fast timescale via suitable techniques of averaging. In Section \ref{sect:averaging} we analyse the feasibility of the averaging method from a dynamical point of view and we exploit it in the vicinity of the upper hyperbolic solution. However, we confirm some limitations of the averaging method: it cannot be applied if the model contains random coefficients generating several ergodic measures and, in any case, the averaged model is not able to reproduce the qualitative properties of the solutions of the nonautonomous model which derive from the temporal variation of the coefficients. 

In Section \ref{sec response sensitivity} the temperature evolution under the nonautonomous EBM and its averaged model is compared by means of a nonautonomous version of the two-point response function of Ashwin and von der Heydt \cite{ashwin2020extreme}. We also study the effect of including a CO$_2$ forcing in the variation of temperature, considering different evolution profiles for the CO$_2$ emissions, such as an instantaneous doubling with respect to the preindustrial stage (see Charney \cite{charney1979}) and the five pathways (SSPs) described by the Intergovernmental Panel for Climate Change (IPCC) \cite{ipcc2023climate}. This is done in terms of the response functions and the new notion of two-point climate sensitivity adapted from the one  in \cite{ashwin2020extreme}. Numerical simulations are run for the quasi-periodic EBM and its version with a chaotic term in the cloud cover coefficient, both with the forcing due to emissions and compared to the preindustrial time. Finally, Section  \ref{secrandom} contains a detailed construction of the chaotic real flow  from which a generic orbit is taken to account for the mentioned chaotic behaviour in the cloud cover.

\section{A nonautonomous zero-dimensional energy balance model}\label{secintro}

In this section, we design a nonautonomous version of a classical energy balance model by introducing a time-dependent variation in the coefficients. Then, we apply techniques of nonautonomous dynamics in order to describe the behaviour of the solutions of the new model.  

\subsection{Fundamental structure of the model}\label{sec:modelsINTRO} We start with a brief introduction to a nonautonomous zero-dimensional energy balance model ($0-$D EBM, for short) of the global climate {\it à la} Budyko-Sellers-Ghil \cite{budyko1969,sellers1969global,ghil1976climate}, considering only coordinate-independent quantities, as in Crafoord and Källén~\cite{crafoord1978} and Ghil and Childress~\cite{ghilchildress1987}. The underlying physical ground is given by an energy
conservation law between incoming
solar radiation $R_{\text{in}}$ and outgoing emission $R_{\text{out}}$: $ c\,T'=R_{\text{in}}-R_{\text{out}}$, where $T(s)$ is the globally averaged surface temperature in Kelvin degrees (i.e.,~$T>0$) of a spherical
planet and the constant $c$
is the thermal inertia of an  ocean mixed layer of depth 30 meters,
covering 70.8\% of the planet \cite{fraedrich1979catastrophes,saltzman2002dynamical,mcguffie2014climate}. The time $s$ is measured in seconds in accordance to the SI for measures and all the constants involved are displayed in Table~\ref{tab:values-parameters}.
 A basic nonautonomous $0-$D EBM 
for the planet's average temperature is 
\begin{equation}\label{eq:Fraed0}
T'(s)=\frac{1}{c}(R_{\text{in}}-R_{\text{out}})=\frac{1}{c}\,\left(\frac{1}{4}\,I(s)\,(1-\alpha(T))-\sigma\,\varepsilon(s,T)\,T^4\right).
\end{equation}
The outgoing radiation $R_{\text{out}}$ is modelled via the Stefan-Boltzmann law with a time-dependent emissivity term with feedback,  
\[
R_{\text{out}}\,=\sigma\,\varepsilon(s,T)\,T^4\,,
\]
where the effective  emissivity $\varepsilon(s,T)$ is obtained as the difference between
the surface emissivity  and the atmospheric emittance, 
and $\sigma$ is the Stefan–Boltzmann constant. 
The incoming solar radiation $R_{\text{in}}$ is modelled as 
\[
 R_{\text{in}}\,= \frac{1}{4}\,I(s)\,(1-\alpha(T))\,,
\]
where $I(s)$
is the total solar irradiance at Earth and $\alpha(T)\in[0,1]$
is the albedo feedback, expressing the fraction of solar radiation which is reflected
from the surface of Earth outside the atmosphere (for example, due to ice, deserts or clouds). The  available data demonstrate that the solar irradiance $I(s)$ varies with time $s\in\R$. Indeed, several modes spanning over timescales going from weeks to thousands of years have been singled out. Further details on the modelling of this aspect are given in Section \ref{subsec:nonautonomous-forcing}, where we present two options for $I(s)$, one quasi-periodic and one almost periodic.


Different parametrisations for the albedo and emissivity feedback processes can be found in the literature (see~\cite{dijdstraViebahn2015,ashwin2020extreme,alexandrovetal2021}, and \cite{widiasih2013} in a latitude-dependent Budyko EBM, among others). We shall fundamentally follow the model by Zaliapin and Ghil \cite{zaliapin2010another} which we present below, although some variants of it are also discussed in Section \ref{sec response sensitivity}.
The albedo feedback is set up as a smooth interpolation between the piecewise-linear formula of Sellers-type models (see~\cite{ghil1976climate,crafoord1978}),  and the piecewise-constant formula of Budyko-type models: 
\[
\alpha(T)= c_1+c_2\frac{1-\mathrm{tanh}\big(k(T-T_c)\big)}{2}\,.
\]
The constants $c_1$ and $c_2$ allow us to fix the asymptotic values of albedo.  It is assumed that at very low temperatures $\alpha(T) \approx 0.85$, whereas at very high temperatures $\alpha(T) \approx 0.15$. The constant $T_c$ is fixed at the freezing point of water, i.e.,   $T_c=273 K$, while the constant $k>0$ allows to change the steepness of the albedo transition depending on the temperature. Values of $k\gg 1$ mimic a Budyko-type model---a discontinuous system in which the albedo takes only two constant values, high and low, depending on whether $T < T_c$ or $T > T_c$. Smaller values of $k$ correspond to Sellers-type models, in which there is a transition ramp between the high and low albedo values. 

The  time- and temperature-dependent emissivity is parametrised following \cite{sellers1969global,crafoord1978,ghilchildress1987,mcguffie2014climate}. We let
\[
\varepsilon(s,T)=1-m(s)\,\tanh{\!\Big((T/T_o)^6\Big)}.  
\]
The function $m(s)$, which accounts for the time dependence of the emissivity, represents the atmospheric opacity. In~\cite{sellers1969global,crafoord1978}  it is taken to be a constant $m=0.5$, in~\cite{zaliapin2010another} $m=0.4$.  We assume that this coefficient varies with time in order to take into account the time-dependent change of cloud cover. Details on the modelling of this aspect will be given in Section \ref{subsec:nonautonomous-forcing}.
The constant $T_o$ is fixed to the normalising value so that $T_o^{-6}=1.9\,{\cdot}\, 10^{-15} K^{-6}$ proposed by Sellers \cite{sellers1969global}. 

Including our choices for the albedo and the greenhouse feedback processes in \eqref{eq:Fraed0} yields the following nonautonomous zero-dimensional energy balance model:
\begin{equation}\label{eq:Ghil}
\begin{split}
    T'(s)=\frac{1}{c}\bigg(\frac{I(s)}{4}\bigg[1-c_1-c_2\frac{1-\mathrm{tanh}\big(k(T-T_c)\big)}{2}\bigg]&\\
    -\sigma\,T^4\bigg[1-m(s)\tanh{\!\Big((T/T_o)^6\Big)}\bigg]&\bigg).
\end{split} 
\end{equation}

\begin{table}
    \centering
    \begin{tabular}{ |p{1.9cm}||p{5.8cm}|p{3.6cm}| }
 \hline
  \multicolumn{3}{|c|}{Constants} \\
 \hline 
 \textbf{Parameter}& \textbf{Meaning} & \textbf{Value according to the SI}\\
 \hline
 $c$ & Constant of thermal inertia & $5{\cdot}10^8\,Jm^{-2} K^{-1}$\\[1ex]
 $\sigma$ & Stefan-Boltzmann constant  &$5.6704{\cdot}10^{-8} \,Wm^{-2} K^{-4}$\\[1ex]
 $c_1$ & Lower-bound for the albedo& 0.15\\[1ex]
$c_2$ & Upper-bound for the increment in albedo& 0.7\\[1ex]
$k$ & Steepness of the albedo transition & $>0$\\[1ex]
 \hline
  \multicolumn{3}{|c|}{Functions} \\
 \hline
 $m(s)$ & High-level IR trapping cloud cover & $[0,1]$\\[1ex]
$I(s)$ & Total solar irradiance at Earth & $[1358.4, 1363.3]\,W m^{-2}$\\[1ex]
 \hline
\end{tabular}
\vspace{1mm}
    \caption{List of constants and functions in the basic model \eqref{eq:Ghil}. 
}
    \label{tab:values-parameters}
\end{table}
For the simulations, we will take the steepness coefficient $k=1$. In addition, we change the timescale from seconds to years, via the change of variable $s= \kappa\, t$, where $\kappa= 60\cdot 60\cdot 24\cdot 365.25$ is the average number of seconds in a year and $t$ is time in years. Then, we obtain the equation
\begin{equation}\label{eq:Ghil-years}
\begin{split}
        T'(t)=\frac{\kappa}{c}\bigg(\frac{I(t)}{4}\bigg[1-c_1-c_2\frac{1-\mathrm{tanh}\big(k(T-T_c)\big)}{2}\bigg]&\\
        -\sigma\,T^4\bigg[1-m(t)\tanh{\!\Big((T/ T_o)^6\Big)}\bigg]&\bigg).
\end{split}
\end{equation}
Note that the functions $\tilde T(t) = T(\kappa\, t)$, $\tilde I(t) = I(\kappa\, t)$ and $\tilde m(t) = m(\kappa\, t)$ have been respectively renamed simply as $T(t)$,  $I(t)$ and  $m(t)$, with a little abuse of notation. Furthermore, we will often refer to \eqref{eq:Ghil-years} as to $T'=g(t,T)$.

\subsection{Modelling the nonautonomous forcing}\label{subsec:nonautonomous-forcing}
In the model \eqref{eq:Ghil-years} there are two time-dependent functions, the total solar irradiance $I(t)$ and the cloud cover $m(t)$. In this paragraph, we discuss how we empirically selected them to replicate some qualitative features of the real phenomena.

\begin{figure}
    \centering
    \begin{overpic}[width=\textwidth]{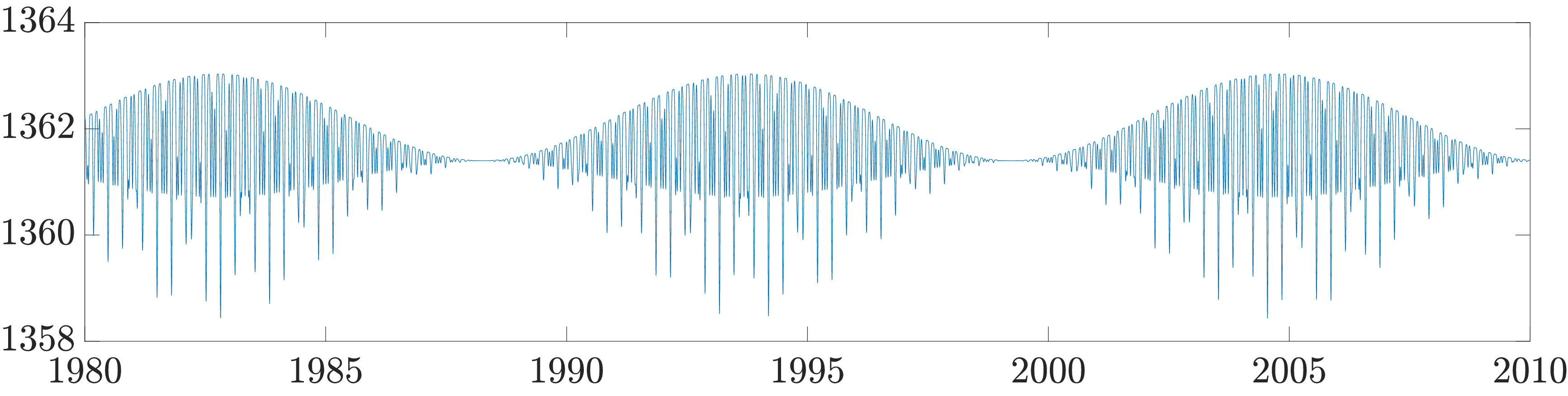}
    \put(-3,9){\begin{turn}{90}TSI ($Wm^{-2}$)\end{turn}}
    \end{overpic}
        \vspace{1mm}
    
    \begin{overpic}[width=\textwidth]{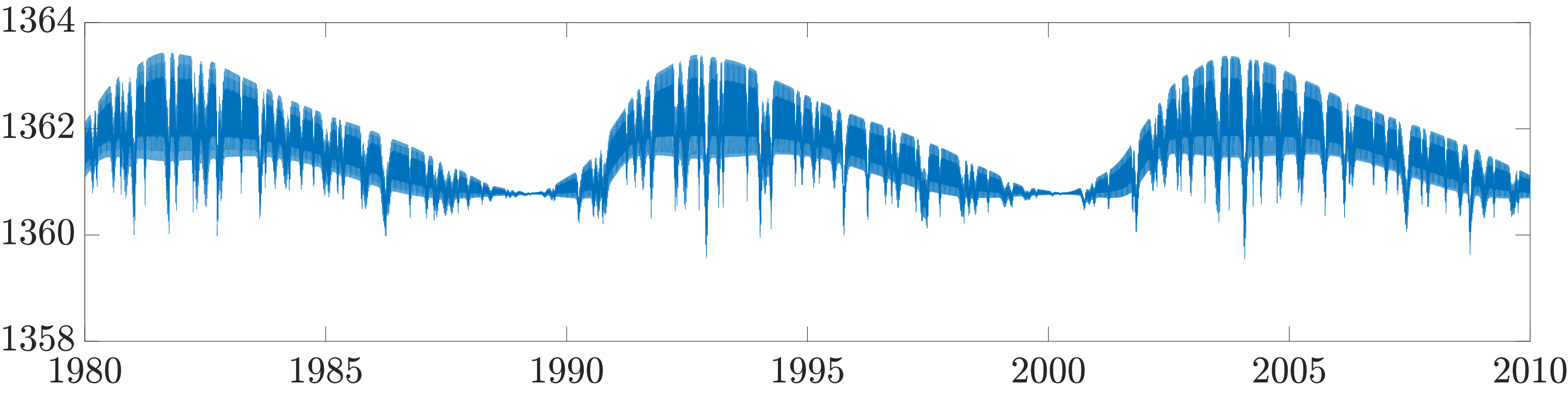}
    \put(-3,9){\begin{turn}{90}TSI ($Wm^{-2}$)\end{turn}}
    \put(43,-3){$t$ (years)}
    \end{overpic}
    \caption{Qualitative shape of the total solar irradiance (TSI) map $I(t)$ for the quasi-periodic model (upper panel) and the almost periodic model (lower panel). For a comparison with an empirical model of the total solar irradiance extrapolated from real data, see Kopp and Lean~\cite{kopp2011new}.}
    \label{fig:TSI}
\end{figure}

The function $I(t)$ captures the following traits of the solar cycles for which satellite data are available~\cite{kopp2011new,schmutz2021changes}: the average low value of total solar irradiance of $I_0= 1360.8\, W m^{-2}$ (recorded during the 2008 solar minimum period), the 11-year cycle of solar activity (also known as the Schwabe cycle) with a maximal increase of total solar irradiance of $1.6 \,W m^{-2}$ ($0.12\%$ of $I_0$), and the decreases of up to $4.6 \, W m^{-2}$ ($0.34\%$ of $I_0$) on time scales ranging from days to weeks.  
We choose the function
\begin{equation*}
\begin{split}
I(t)=&1360.8\, \Bigg[1+\frac{0.0012}{2}\left(1+\sin\! \bigg(\frac{2\pi t}{11}\bigg)\right)\\
&-\frac{0.0034}{4}\left(\sin^{10} \!\big(\sqrt{97}\pi t\big)+\sin^{10} \!\big(\sqrt{47}\pi t\big)\right)\left(1+\sin\! \bigg(\frac{2\pi t}{11}\bigg)\right)\!\Bigg].
\end{split}
\end{equation*}
We note that the function $I(t)$ has maximum equal to $ 1363.3\,Wm^{-2}$, minimum equal to $ 1358.4\,Wm^{-2}$  and numerically estimated average over an interval of time of five thousand years equal to $1361.6\,Wm^{-2}$. An alternative approach (which is however costly and hard to implement without showing appreciable differences in the numerical experiments) is to consider an almost periodic forcing of the type
\[
\begin{split}
    I_{ap}(t)= I_0 +\frac{1.6}{2}\Bigg(1+CL_2\bigg(\frac{2\pi t}{11}\bigg)\Bigg)+\frac{\phi(t)}{2}\Bigg(1.1+CL_2\bigg(\frac{2\pi t}{11}\bigg)\Bigg), \quad\text{where}\\
    CL_2(\theta)=\sum_{j=1}^{\infty}\frac{\sin(j \theta)}{j^2}\quad(\text{Fourier series of Clausen function $CL_2$)}\ \quad\text{and}\\
    \phi(t)=-\frac{4.6}{5}\!\sum_{j=1,3,5,7,11}\!\!\!\!\!\sin^{180}\!\bigg(\frac{2\pi t}{\sqrt{j}}\bigg)+\frac{1}{2}\big(\sin^3(52.18\cdot 2\pi t)+\sin^3(365.25\cdot 2\pi t)\big).
\end{split}
\]
The rescaled Clausen function simulates the baseline 11-year cycle of solar activity, while the function $\phi(t)$ accounts for changes at lower timescales. In Figure~\ref{fig:TSI}, it is possible to appreciate the qualitative shape of $I(t)$ and $I_{ap}(t)$. For a comparison with an empirical model of the total solar irradiance extrapolated from real data, see Kopp and Lean~\cite{kopp2011new}.

 \par \smallskip

\begin{figure}
    \centering
    \begin{overpic}[width=0.45\textwidth]{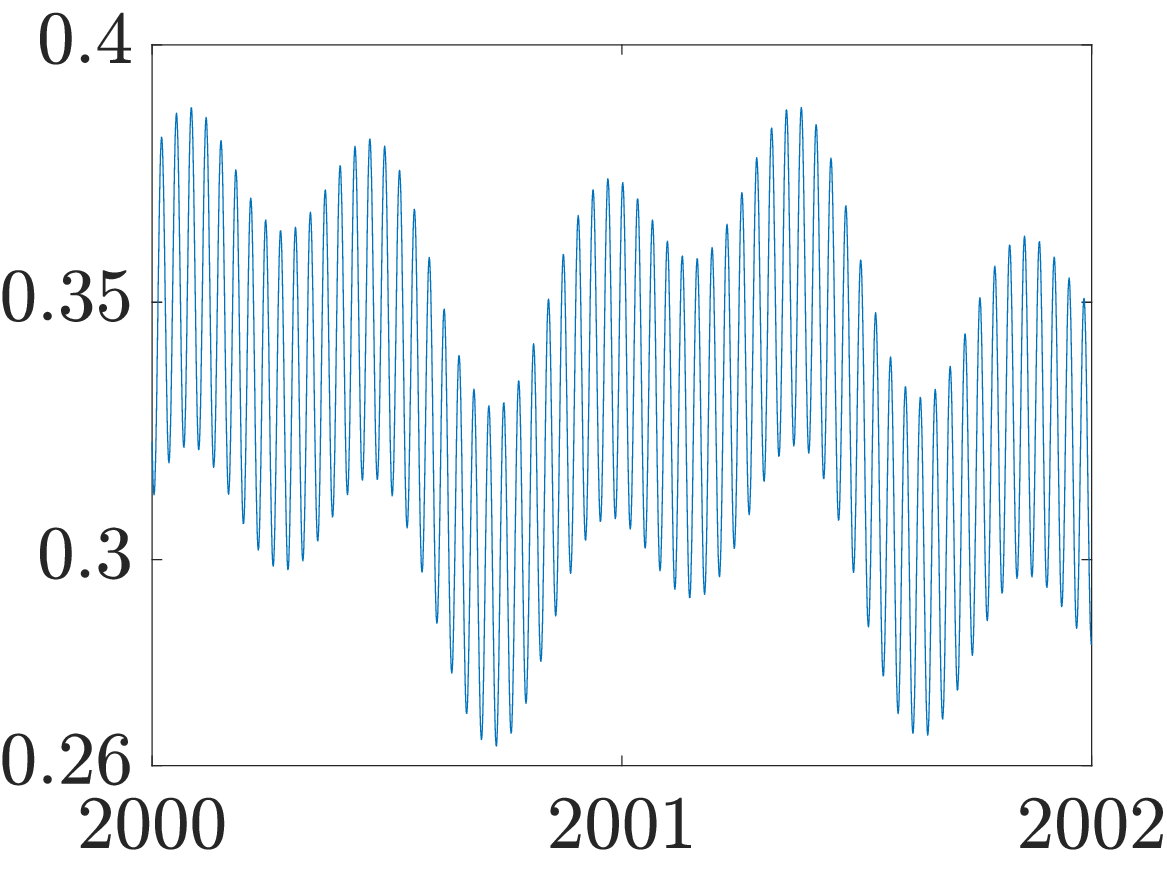}
    \put(43,-4){$t$ (years)}
    \end{overpic}
    \hspace{0.5cm}
    \begin{overpic}[width=0.45\textwidth]{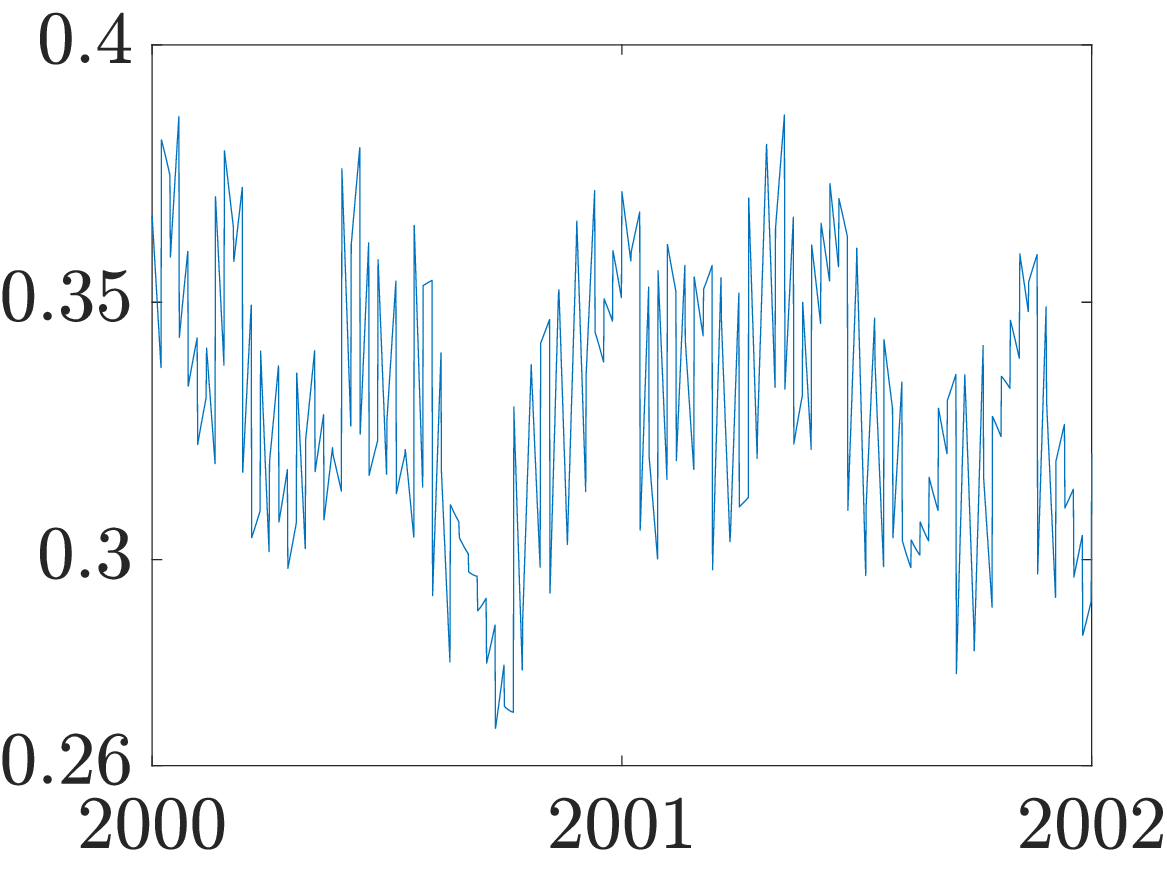}
    \put(43,-4){$t$ (years)}
    \end{overpic}
    \caption{Modelling of thin clouds in the upper layer of the atmosphere responsible for the reflection of most of long-wave radiation from the surface. On the left-hand side a quasi-periodic model is shown, whereas on the right-hand side the quasi-periodic forcing is perturbed by a chaotic dynamical system.}
    \label{fig:clouds}
\end{figure}

In regard to the atmospheric opacity $m(t)$, we aim to capture changes of the \emph{effective cloud amount}, that is, the cloud amount
weighted by cloud emissivity,  which is responsible for most of the long-wave radiation reflected back to the planet. According to \cite{stubenrauch2013assessment}, the effective cloud amount has a global average of about 50\%. In order to balance the model around the current average temperature of $288.5\, K$, we choose an average of 33\%.  
To introduce a time-dependent variability of the effective cloud amount, we include four different timescales and respective amplitudes: 7--14 days with a 0.1 anomaly, 5--6 months and one-year seasonal effects with a cumulative 0.1 anomaly, and interannual variability (5-7 years) akin to El Niño with a 0.01 anomaly~\cite{stubenrauch2013assessment}, 
\[
m(t)=0.33\,\big(1+0.1\, m_d(t)+0.05\,(m_h(t)+m_y(t))+0.01\, m_{N}(t) \big)\,.
\]
We consider two approaches. The first option  is a quasi-periodic function,
\[
\begin{split}
    m_1(t)=0.33\bigg(1&+0.1\, \sin\!\big(2\sqrt{997}\pi t\big)\\&+ 0.05\,\big( \sin\!\big(2\sqrt{5}\pi t\big)+\sin(2\pi t)\big)+0.01\, \sin\!\bigg(\frac{2\pi t}{\sqrt{29}}\bigg)\! \bigg).
\end{split}
\]
The second one is a forcing with ingredients of chaotic dynamics, trying to account for the intrinsic unpredictability in the cloud cover (see~\cite{liuetal2023cloudcoverage}), 
\begin{equation}\label{eq:m2}
\begin{split}
    m_2(t)=0.33\bigg(1&+0.1\, {\rm Im}(p_\Theta(t))\\
    &+ 0.05\,\big( \sin\!\big(2\sqrt{5}\pi t\big)+\sin(2\pi t)\big)+0.01\, \sin\!\bigg(\frac{2\pi t}{\sqrt{29}}\bigg)\! \bigg),  
\end{split}  
\end{equation}
where ${\rm Im}(p_\Theta(t))$ corresponds to the imaginary part of the evaluation at $0$ of an orbit inside a chaotic set of functions built upon linear interpolation of orbits for the expanding map $\theta\mapsto 2\theta \!\mod 2\pi$ on the unit circle. The detailed construction has been included in Section~\ref{secrandom}. For now, we refer to Figure~\ref{fig:clouds} to appreciate the difference between the two types of forcing. In the simulation, we have taken the semiorbit of the angle $\pi\sqrt 5/2 \!\!\mod 2\pi$ under the doubling map. 

\subsection{Existence of exponentially stable and unstable solutions}\label{subsec:attr-repel-pair}

In this section we aim to characterize the possible dynamical scenarios and bifurcations for \eqref{eq:Ghil-years} by determining its regions of concavity and convexity and studying the sign of the vector field. We will be supported by recent results for scalar nonautonomous differential equations with concave and coercive nonlinearities by Longo et al.~\cite{longo2024critical}, which, to a considerable extent, rely on an underlying skew-product flow (or nonautonomous dynamical system), which we briefly introduce below. 

Dynamical methods applied to ODEs, PDEs or stochastic equations modellling different climate issues have  been used in the last decades, see~\cite{dijkstra2005Book,ghiletal2008climatedynamics,chekrounetal2011stochasticclimate,DeSaedeleeretal2013,widiasih2013, ashwin2020extreme,alexandrovetal2021}, among others. 
Particularly, Ghil et al.~\cite{ghiletal2008climatedynamics} suggested the use of random dynamical systems when there is a noise forcing modelled by a stationary stochastic process. 
Random dynamical systems are structurally close to skew-product flows since the dynamical system is defined on a product space $\pi:\R\times\Omega\times X\to \Omega\times X$, $(t,\omega,x)\mapsto (\theta_t(\omega),\varphi(t,\omega,x))$ where the first component is a measurable flow on a probability space $\Omega$, which is measure preserving, and the second component is given by a cocycle map: $\varphi(t_1+t_2,\omega,x)=\varphi(t_1,\theta_{t_2}(\omega),\varphi(t_2,\omega,x))$, in order to have the standard dynamical system conditions $\pi_0=\rm{id}$ and $\pi_{t_1+t_2}=\pi_{t_1}\circ \pi_{t_2}$ for all $t_1,t_2\in\R$, where $\pi_t(\omega,x)=\pi(t,\omega,x)$, $\rm{id}$ is the identity map on $\Omega\times X$, and $\circ$ is the composition operator.  A skew-product flow has the same structure, except for the fact that the base space $\Omega$ is metric and compact and the flow $\theta$ is continuous.  By the Krylov-Bogoliubov theorem \cite{kryloff1937theorie}, the set of $\theta$-invariant measures on $\Omega$ is nonempty. Thus, ergodic theory can be applied in both contexts. Also, the dynamical system is measurable or continuous, depending on the measurable and/or topological properties of both the phase space $X$ and the cocycle map $\varphi$. If the map $\pi$ is defined only for nonnegative times $t\geq 0$, it is called a random semi-dynamical system or a skew-product semiflow, respectively.

Let us explain how we fall into the skew-product flows context. It is well-known that the solutions of a scalar nonautonomous differential equation 
such as $T'=g(t,T)$ in \eqref{eq:Ghil-years} do not induce a dynamical system in a direct way. Nevertheless, one observes that the time-translation of a solution $T_\tau(t)=T(t+\tau)$ is a solution of the translated equation   $T'=g_\tau(t,T)=g(t+\tau,T)$. Hence,  it is natural  to consider the set of translated maps $\{g_\tau\mid\tau\in\R\}\subset Y$ and, depending on the properties of $g$, look for a suitable space of functions $Y$ and a topology so that the  topological closure of the set of time-translations, i.e.,~the  so-called hull of $g$, 
\begin{equation}\label{eq:hull}
\mathcal{H}=\rm{cls}\{g_\tau\mid\tau\in\R\}\,,    
\end{equation}
is a compact metric space and the shift flow $\theta:\R\times \mathcal{H}\to \mathcal{H}$, $(t,h)\mapsto h_t$ is continuous. In this case, one says that the time variation in the equation can be compactified. Here we assume that $g(t,T)$ is bounded and uniformly continuous on $\R\times K$ for compact sets $K\subset\R$, as then the hull is compact for the compact-open topology on the space of real continuous maps on $\R^2$.  
Note that with any of the terms $I(t)$ and $m(t)$ considered in Section~\ref{subsec:nonautonomous-forcing}, $g$ satisfies this condition.  So, let $\mathcal{H}$ be the hull of $g$, and for each $h\in\mathcal{H}$, denote by $u(t,h,T_0)$ the solution of the equation $T'=h(t,T)$ with initial condition $u(0,h,T_0)=T_0$. Then, we can define a continuous skew-product flow 
\begin{equation*}
 \begin{array}{rccl}
 \pi: &\mathcal{U}\subseteq \R\times\mathcal{H}  \times \R^+ & \longrightarrow & \mathcal{H}  \times \R^+ \\
 &(t,h,T_0) & \mapsto &(h_t,u(t,h,T_0))\,,
\end{array}
\end{equation*}
on an appropriate open set $\mathcal{U}$ subject to the existence of the solutions. Since the equations are scalar, this flow is monotone, that is, the order of initial data is preserved along the trajectories.  


At this point, let us focus on the study of the regions of concavity and convexity for the map $g(t,T)$ with respect to $T$. To this end, we calculate the partial derivatives up to order two of the right-hand side of \eqref{eq:Ghil-years} with respect to the variable $T$:
\[
\begin{split}
  \frac{\partial g}{\partial\, T}(t,T)&= \frac{\kappa}{c} \bigg( \frac{c_2\, k\, I(t)}{8}\sech^2\!\big(k(T-T_c)\big)-4\,\sigma\, T^3\big(1-m(t)\tanh\!\big((T/ T_o)^6\big)\big)\\ 
  &\qquad\qquad+\frac{6\,\sigma\, m(t)\,T^9}{ T_o^6}\sech^2\!\big((T/ T_o)^6\big)\! \bigg)
,\\
  \frac{\partial^2 g}{\partial\, T^2}(t,T)&=\frac{\kappa}{c} \bigg(-\frac{c_2\, k^2\, I(t)}{4}\tanh\!\big(k(T-T_c)\big)\sech^2\!\big(k(T-T_c)\big)+ 
  \\
  &\qquad\quad \, \, -12\,\sigma\, T^2\big(1-m(t)\tanh\!\big((T/ T_o)^6\big)\big)+\\
  &\qquad\quad \, \,+\frac{78\,\sigma\, m(t)\,T^8}{ T_o^6}\sech^2\!\big((T/ T_o)^6\big)+\\ 
  &\qquad\quad \, \, -\frac{72\,\sigma\, m(t)\,T^{14}}{ T_o^{12}}\tanh\!\big((T/ T_o)^6\big)\sech^2\!\big((T/ T_o)^6\big)\!\bigg).\\
\end{split}
\]
The right-hand side of \eqref{eq:Ghil-years} is convex where its second derivative with respect to $T$ is positive for all $t\in\R$, and concave where it is negative for all $t\in\R$.  Therefore, we need to study the sign of $({\partial^2 g}/{\partial\, T^2})(t,T)$. 
This is not immediately obvious as the previous inequality is not algebraically solvable in an easy way. Nonetheless, we can numerically explore the sign of $\partial^2 g/\partial\,T^2$, which will be mostly sufficient for our objective. 

\begin{figure}[h]
    \centering
    \begin{overpic}[width=\textwidth]{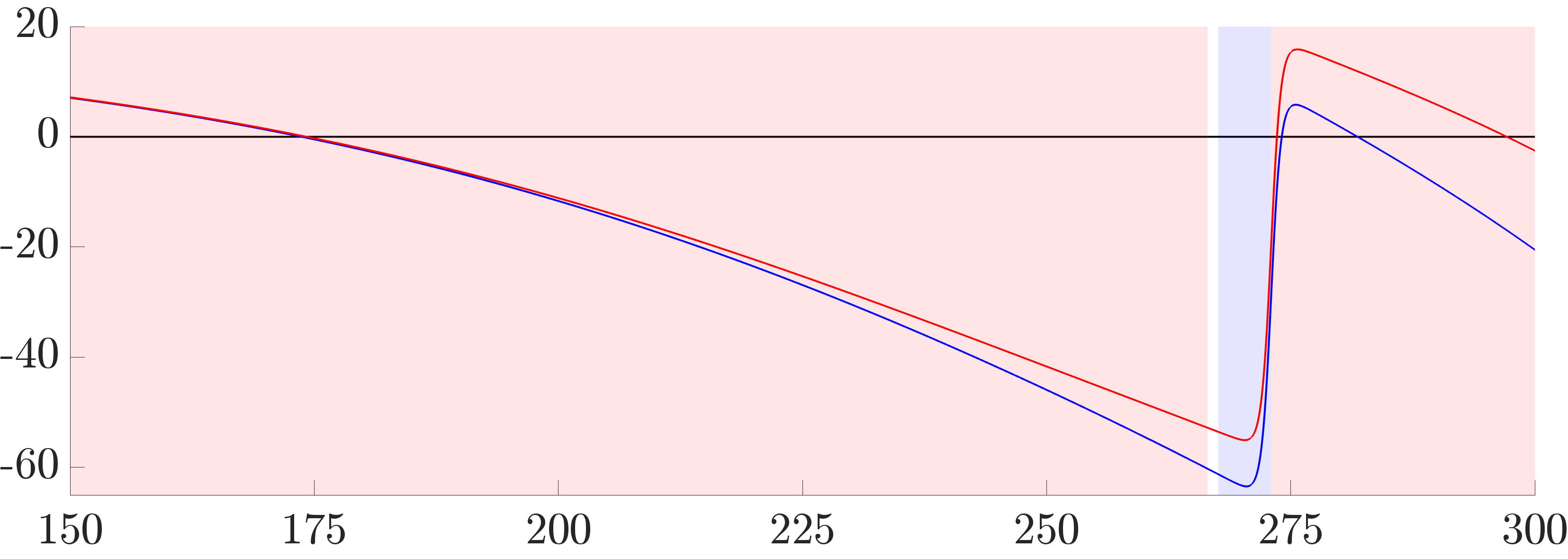}
    \put(48,-2){$T\, (K)$}
    \end{overpic}
    \caption{Graphs of  $g_*(T)$ (blue solid line) and $g^*(T)$ (red solid line) with respect to the variable $T$. The coloured areas correspond to values of $T$ where the sign of the second derivative of $g$ with respect to $T$  is positive (blue shaded region) or negative (red shaded regions) for all values of $t$. It is possible to appreciate that the vector field $g$ remains negative for all values of $T\in[176,273.5]$, thus preventing the existence of an attractor-repeller pair in the convex region.}
    \label{fig:conc-conv}
\end{figure}

A numerical estimation of the variation of the coefficients in the quasi-periodic model yields the values $1358.4 \leq I(t)\leq 1363.3$ and $0.2609\leq m(t)\leq 0.3988$, $t\in\R$. The following conclusions apply to \eqref{eq:Ghil-years} as far as  $g_*(T)\leq g(t,T)\leq g^*(T)$, $t\in\R$ where the functions $I(t)$ and $m(t)$ are kept constant at their maximum in $g^*(T)$ (red solid line in Figure~\ref{fig:conc-conv}) or at their minimum in $g_*(T)$ (blue solid line in Figure~\ref{fig:conc-conv}). The coloured regions in Figure \ref{fig:conc-conv} correspond to the parts of the phase space where $g$ is concave (in red) and convex (in blue) uniformly on $t\in\R$. 
As far as we are concerned, we appreciate that the intervals $[0, 266]$ and $[273.5, 305]$ are in the concave regions of the phase space, and that the interval $[268,272.5]$ is in the convex one.

\begin{figure}
    \centering
    \begin{overpic}[width=0.45\textwidth]{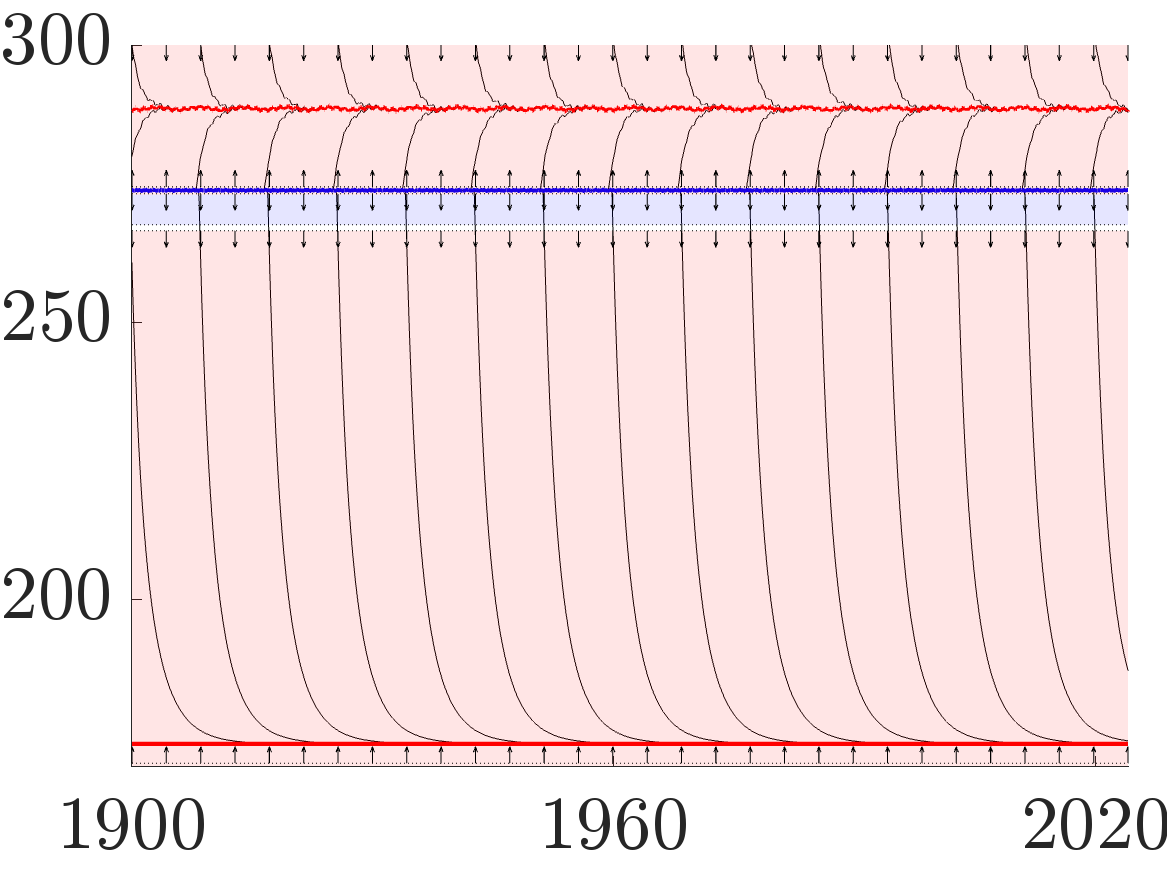}
    \put(43,-4){$t$ (years)}
    \put(-6,35){\begin{turn}{90} $T\, (K)$ \end{turn}}
    \end{overpic}
    \hspace{0.5cm}
    \begin{overpic}[width=0.45\textwidth]{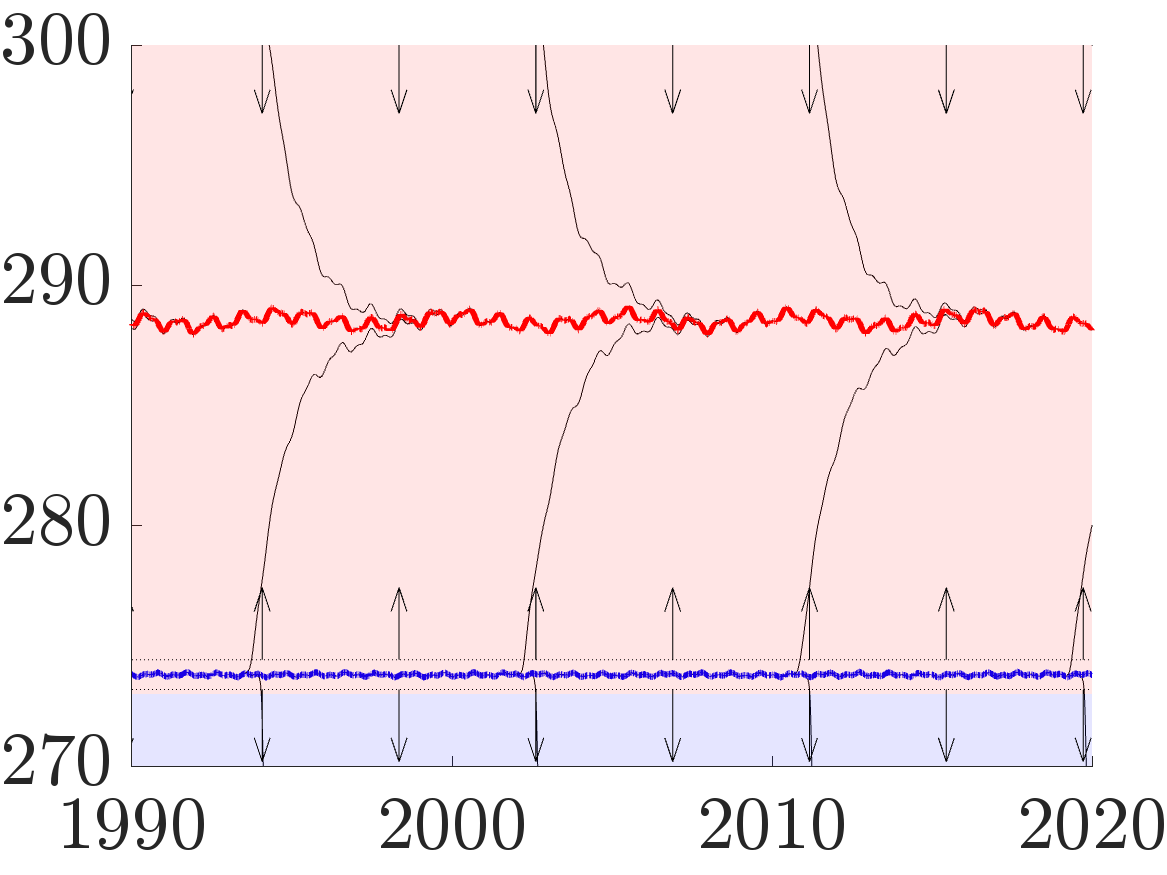}
    \put(43,-4){$t$ (years)}
    \put(-6,35){\begin{turn}{90} $T\, (K)$ \end{turn}}
    \end{overpic}
    \caption{Attractors (red curves), repeller (blue curve), sub- and super-equilibria and zones of concavity (red shaded regions) and convexity (blue shaded regions) for the quasi-periodic model.}
    \label{fig:semiequilibria}
\end{figure}

We proceed to describe the dynamics of the solutions of the EBM \eqref{eq:Ghil-years}. There exists a so-called {\it attractor-repeller pair} of  bounded solutions $a_1(t)$, $r(t)$, $t\in\R$ and a second attracting solution $a_2(t)$, $t\in\R$ such that $a_2(t)<r(t)<a_1(t)$ for all $t\in\R$, and if $T(t,t_0,T_0)$ denotes the solution of $T'=g(t,T)$ with $T(t_0,t_0,T_0)=T_0$, then there are constants $\rho,\rho_1, \rho_2>0$ and for every $\nu>0$ there exist constants $k_\nu, k_{1,\nu}, k_{2,\nu}>1$ such that: 
\begin{equation}\label{eq: atractor a_1}
\begin{split}
    |a_1(t)-T(t,t_0,T_0)|\leq & k_{1,\nu}\,e^{-\rho_1 (t-t_0)}\, |a_1(t_0)-T_0|\\ &\qquad\quad\hbox{if}\;\,r(t_0)+\nu \leq T_0\leq 305 \,\;\hbox{and}\;\, t\geq t_0\,,\\
|r(t)-T(t,t_0,T_0)|\leq &k_\nu\,e^{\rho (t-t_0)}\, |r(t_0)-T_0|\\ &\qquad\quad\hbox{if}\;\, 273.5 \leq T_0\leq a_1(t_0)-\nu  \,\;\hbox{and}\;\, t\leq t_0\,,\\
|a_2(t)-T(t,t_0,T_0)|\leq &k_{2,\nu}\,e^{-\rho_2 (t-t_0)}\, |a_2(t_0)-T_0|\\ &\qquad\quad\hbox{if}\;\, \nu \leq T_0\leq 266 \,\;\hbox{and}\;\,t\geq t_0\,.
\end{split}
\end{equation}
See Figure \ref{fig:semiequilibria} for some numerical computations of the attractors and repellers for the quasi-periodic model and let us mathematically justify the previous assertion and describe the global dynamics following a series of steps.

{\it Step 1.} In the temperature band $[273.5,305]$ where the equation is strictly concave, there is an attractor-repeller pair of hyperbolic solutions $a_1(t)$, $r(t)$, $t\in\R$ that behave as indicated in \eqref{eq: atractor a_1}. 

 In order to provide a rigorous proof of the existence of the attractor-repeller pair, we use the skew-product formalism. First, since $g_*(T)\leq g(t,T)\leq g^*(T)$ for all $t\in\R$, then, we also have that  $g_*(T)\leq h(t,T)\leq g^*(T)$ for all $t\in\R$ and $h\in\mathcal{H}$ thanks to the hull construction and the continuity of the flow. Moreover, since $ 0<g_*(275.5)\leq h(t,275.5)$ and $h(t,305)\leq g^*(305)<0$ for all  $t\in\R$ and  $h\in\mathcal{H}$, the maps $a_0,b_0:\mathcal{H}\to \R$, $a_0\equiv 275.5$ and $b_0\equiv 305$ are a continuous sub-equilibrium and a continuous super-equilibrium, respectively, for the skew-product semiflow $\pi$ (see Proposition 4.4 in Novo et al.~ \cite{novo2005almost}). In this situation, there exists a semicontinuous equilibrium $A_1:\mathcal{H} \to [275.5,305]$, that is,  $A_1(h_t)=u(t,h,A_1(h))$ for all $h\in \mathcal{H}$ and $t\in\R$ (see Theorem 3.6 in \cite{novo2005almost}).  In particular $a_1(t):=A_1(g_t)$, $t\in \R$ is an entire bounded solution of the EBM. 
Second, since $g^*(273.5)<0$ and $g_*(275)>0$,   in the negatively invariant band of temperatures $[273.5,275]$ we can reverse the time $\widetilde u(t,h,T_0)= u(-t,h,T_0)$ and consider the skew-product semiflow $(t,h,T_0)\mapsto (h_{-t},\widetilde u(t,h,T_0))$ to fall into a situation similar to the one before. Reversing time back, we obtain the existence of a second equilibrium
$R:\mathcal{H} \to [273.5,275]$ such that $R(h_t)=u(t,h,R(h))$ for all $h\in \mathcal{H}$ and $t\in\R$.  Then, $r(t):=R(g_t)$, $t\in \R$ is another entire bounded solution of the EBM. Furthermore, by construction both solutions are uniformly separated. If we cut the field $g(t,T)$ at the temperature $273.5$ and extend it below this value with  the second order Taylor expansion of $g(t,T)$ at $273.5$, we obtain a strictly concave and coercive scalar equation that has two uniformly separated bounded solutions. In this situation, Theorem 3.4 in \cite{longo2024critical} says that both solutions are hyperbolic, forming an attractor-repeller pair which, in particular,  fulfills \eqref{eq: atractor a_1}.  

{\it Step 2.} In the positively invariant band of temperatures $[0,266]$ where the equation is strictly concave, there is an attracting  hyperbolic solution $a_2(t)$, $t\in\R$, which behaves as indicated in \eqref{eq: atractor a_1}.

A suitable study of the sign of the vector field, just as before, allows us to  deduce the existence of an equilibrium $A_2:\mathcal{H} \to [1,266]$,   $A_2(h_t)=u(t,h,A_2(h))$ for all $h\in \mathcal{H}$ and $t\in\R$.  In particular $a_2(t):=A_2(g_t)$, $t\in \R$ is an entire bounded solution of the EBM. Furthermore, we can choose a negative $T_1<0$ such that $g^*(T_1)<0$ and then argue in the negatively invariant band $[T_1,0]$ as in Step 1, to get a second bounded solution uniformly separated from $a_2(t)$. This solution has no physical meaning for the EBM, but it permits us to apply Theorem 3.4 in \cite{longo2024critical}---once more  after cutting the field at the temperature $266$ and using a second order Taylor expansion at $266$ to extend it above this value, so as to have a strictly concave and coercive vector field everywhere---to deduce that $a_2(t)$ is a hyperbolic attracting solution.  
 
{\it Step 3.}  We now consider the zone between the regions in Steps 1 and 2, that is, the interval of temperatures $[266,273.5]$. Since the field $g(t,T)\leq g^*(T) < 0$, for all $T\in[266,273.5]$ and $t\in\R$, then the corresponding flow is pointing downwards. Therefore,  $[266,273.5]$ is to be seen as a transit zone for the solutions traveling downwards until they reach the inferior limit of $T=266$ and then, they get attracted by the solution $a_2(t)$.

The above arguments provide a rigorous and complete description of the behaviour of solutions of the EBM for all initial conditions $T_0\in [0,305]$.
See  Figure \ref{fig:semiequilibria} for a global picture of the dynamics. We finish this section with two comments. First, it is worth mentioning that the proof of the above-cited result \cite[Theorem 3.4]{longo2024critical} relies on an adaptation of the dynamical methods developed in Longo et al.~\cite{longo2021rate}. In particular, it ensues that the maps $A_1$, $A_2$ and $R$ are continuous on $\mathcal{H}$ (see the proof of \cite[Theorem 3.5]{longo2021rate}). Thus, the solutions $a_1(t)=A_1(g_t)$, $a_2(t)=A_2(g_t)$ and $r(t)=R(g_t)$  reproduce the dynamics of the hull, which is determined by the temporal variation of the vector field $g(t,T)$. For instance, in the quasi-periodic EBM, the hull $\mathcal{H}$ of $g$ is a torus, and then, also the maps $a_1(t)$, $a_2(t)$ and $r(t)$, given by the evaluation of a continuous map along an orbit in the torus, are all quasi-periodic. 

Finally, the standard property of robustness of hyperbolicity for differential equations (e.g., see \cite[Proposition 3.3]{longo2024critical}) affirms that given an $\ep>0$ there exists a $\delta=\delta(\ep)>0$ such that, if the equation with a hyperbolic solution  is perturbed by a term of norm up to $\delta$, then the perturbed equation also has a hyperbolic solution which is located within distance $\ep$ of the initial hyperbolic solution.  The map $\delta(\ep)$ can be seen as an indicator of climate sensitivity  and we will come back to this important fact in Section \ref{secrandom}. Let us mention here that in an autonomous setting, a new gauge called {\it intensity of attraction}, which measures the robustness of attractors in metric terms,  has recently been introduced in Meyer and McGehee~\cite{meyermcgehee2022}. 

\section{The averaging method and its limitations}\label{sect:averaging}
Energy balance models of the form \eqref{eq:Ghil} are frequently analysed as autonomous systems (see~\cite{north1981energy}-\cite{Chao2021assessment}, \cite{budyko1969,sellers1969global,ghil1976climate}). This naturally raises the question of how such an approximation is justified, given that the underlying system—namely, the Earth's climate—is inherently subject to time-dependent forcing. A common justification lies in the possibility of temporal averaging, supported by the observation that many relevant external influences, such as astronomical and meteorological factors, exhibit well-defined characteristic frequencies. 

However, a rigorous justification for the use of averaging to derive autonomous energy balance models is lacking in the existing literature. In this section, we analyse the conditions under which it is valid to approximate an EBM as autonomous, based on the error bounds provided by the averaging method, and identify scenarios in which such an approximation is not warranted.

\subsection{Introduction to averaging}\label{subsec:intro averaging}
Consider a function $f:[0,\infty]\times\R\to\R$, $(t,T)\mapsto f(t,T)$ bounded, continuous, and locally Lipschitz continuous with respect to the variable $T$, i.e.,~such that for every $r>0$ there exists a constant $L_r>0$ such that 
\[
|f(t,T_1)-f(t,T_2)|\le L_r\,|T_1-T_2|\quad \text{for all } t\geq 0 \,\text{ and }\, |T_1|,|T_2|\le r\,.
\]
The function $f$ is called a KBM function---after the mathematicians Krylov, Bogoliubov and Mitropolsky---if the average
\begin{equation}\label{eq:average-0}
\widehat f(T)=\lim_{R\to\infty} \frac{1}{R}\int_0^Rf(t,T)\,dt
\end{equation}
exists for every $T\in\R$. It is easy to check that, with the local Lipschitz property, this limit is uniform for $T$ in compact sets. The theory of averaging (see Bogoliubov and Mitropolsky \cite{bogoliubov1961asymptotic} and Sanders et al.~\cite{sanders2007averaging}) aims at establishing relations between the solutions of the nonautonomous Cauchy problem
\begin{equation}\label{eq:23/01-16:45}
T'\!= \ep\,f(t,T)\,,\quad  T(0)=T_0\,,
\end{equation}
and the autonomous averaged Cauchy problem 
\begin{equation}\label{eq:23/01-16:46}
     T'\!=\ep\,\widehat f(T)\,,\quad  T(0)=T_0\,,
\end{equation}
where $T_0\in\R$ and $\ep>0$ is a small number. In fact, under certain conditions, it is possible to establish an upper-bound to the distance between bounded solutions $T_\ep(t)$ of \eqref{eq:23/01-16:45} and $T_\ep^*(t)$ of \eqref{eq:23/01-16:46} in long time intervals $[0,1/\ep]$. This is usually written, using the standard Landau's notation, as
\[
\sup_{t\in[0,1/\ep]}|T_\ep(t)- T_\ep^*(t)| \sim O(\delta_n(\ep)^\alpha) 
\]
for some $\alpha>0$, where $\delta_n$ is the so-called convergence function defined by 

\begin{equation}\label{eq:convergence-function-delta}
 \delta_n(\ep)=\sup_{|T|\le n}\sup_{0\le \ep R\le 1}\ep\,\bigg|\int_0^R\big(f(t,T)-\widehat f(T)\big)\, dt\bigg|\,,\quad n\in\N\,.   
\end{equation}
Note that, if $f$ is KBM, then $\lim_{\ep\to0^+}\delta_n(\ep)=0$ for every $n\in\N$, and conversely, if for some real numbers $\widehat f(T)$, $T\in\R$ the previous limits are null, then $f$ is KBM with averages $\widehat f(T)$, $T\in\R$. A stronger assumption is that the limit in \eqref{eq:average-0} does not depend on the initial point of the interval of integration. We say that $f$ is uniformly KBM (UKBM, for short) if the following limit exists for every $T\in\R$, 
\begin{equation}\label{eq:average-t0}
\widehat f(T)=\lim_{R\to\infty} \frac{1}{R}\int_{t_0}^{t_0+R}f(t,T)\,dt\,,\quad \text{uniformly in }t_0\ge 0\,.  
\end{equation}
It is easy to check that $f$ is UKBM if and only if $\lim_{\ep\to0^+}\widetilde\delta_n(\ep)=0$ for every $n\in\N$, where
\begin{equation}\label{eq:convergence-function-delta-hat}
\widetilde\delta_n(\ep)=\sup_{|T|\le n}\!\sup_{\substack{t_0\ge0\\[1.5pt] 0\le \ep R\le 1}}\ep\,\bigg|\int_{t_0}^{t_0+R}\big(f(t,T)-\widehat f(T)\big)\, dt\bigg|\,.
\end{equation}

\subsection{Application to the energy balance model} \label{subsec:averaged-EBM} 
The  theory of averaging offers quantitative bounds for the error  in terms of the small parameter $\ep$ in the equations \eqref{eq:23/01-16:45}-\eqref{eq:23/01-16:46}, but the results derived apply to each $\ep$ separately. In order to benefit from this fact, we rewrite equation~\eqref{eq:Ghil-years} taking out a factor $\sqrt[3]{\sigma}$, so that the EBM shows a structure compatible with the theory of averaging, namely,
\begin{equation}\label{eq:Ghil-years-ep}
\begin{split}
     T'=\varepsilon_0\bigg(\frac{I(t)}{4\sqrt[3]{\sigma}}\bigg[1-c_1-c_2\frac{1-\mathrm{tanh}\big(k(T-T_c)\big)}{2}\bigg]&\\
     -\sqrt[3]{\sigma^2}\,T^4\bigg[1-m(t)\tanh{\!\Big((T/ T_o)^6\Big)}\!\bigg]&\bigg),
\end{split}
\end{equation}
where $\varepsilon_0=\kappa\sqrt[3]{\sigma}/c\approx1.2\,{\cdot}\,10^{-3}$. We will refer to \eqref{eq:Ghil-years-ep} as to $T'=\varepsilon_0 f(t,T)$. Note that the dependence on time of this model is completely encapsulated in the functions $I(t)$ and $m(t)$,  which are assumed to have averages $\widehat I$ and $\widehat m$, respectively, from now on. Note that this is the case with quasi-periodic or almost periodic coefficients, but it is not expected to be the case when $m(t)$ incorporates a chaotic variation---see Section \ref{subsec:nonautonomous-forcing}. Specifically, the averaged equation is  written as 
\begin{equation}\label{eq:averaged EBM}
\begin{split}
    T'=\ep_0\bigg(\frac{\widehat I}{4\sqrt[3]{\sigma}}\bigg[1-c_1-c_2\frac{1\!-\!\mathrm{tanh}\big(k(T-T_c)\big)}{2}\bigg]&\\
    -\sqrt[3]{\sigma^2}\,T^4\bigg[1-\widehat m\,\tanh{\!\Big((T/ T_o)^6\Big)}\bigg]&\bigg).
\end{split}
\end{equation}

The time-varying map $f$ has a reasonable size, which is in line with the spirit of the averaging method. 
Also, note that one can consider a slow timescale $\tilde t$ measured in 1000 years, so that the relation between the fast timescale $t$ in years and the slow one is   $\tilde t = 10^{-3} \,t$, which is approximately $\tilde t = \varepsilon_0\, t$ and the model is expected to yield information valid for around 300 years. 

It should be noted, however, that the results of averaging are typically written either for bounded and globally Lipschitz functions, or restricted to a bounded ball centered at the origin for the spatial variable. 
It is clear that, under the assumption of local Lipschitz continuity in $T$, one can pass from the latter case to the former by using a suitable smooth cut-off function, also called a mollifier, i.e.,~a $C^\infty$ function $\chi:\R\to[0,1]$ such that $\chi(T)=1$ for $T\in[100,400]$ and $\chi(T)=0$ for $T\le 0$ and $T\ge 500$. The interval $[100,400]$ is chosen to guarantee that for physically meaningful temperatures the mollified model coincides with the original one. The problem $T'=\varepsilon_0 f(t,T)\,\chi(T)$ is globally Lipschitz continuous in $T$, bounded and maintains the same solutions of the original energy balance model in the region of the phase space that is physically meaningful. 
For the sake of simplicity, we omit this technical adjustment in the next sections.

\subsection{A dynamical understanding of the existence of averages}\label{subsec:dynamical underst}
The limits in \eqref{eq:average-0} and \eqref{eq:average-t0} admit a dynamical interpretation thanks to  Bebutov's formalism and Birkhoff's ergodic theorem, as we clarify in this section. As noticed in the previous section, the vector field $f$ in \eqref{eq:Ghil-years-ep} is KBM (resp.~UKBM) if and only if both $I(t)$ and $m(t)$ are KBM (resp.~UKBM). For this reason, in relation to the behaviour of the averages, we can restrict our analysis to the functions $I(t)$ and $m(t)$. So, let us just consider a real map $p:\R\to \R$.

Let us start by taking a bounded and continuous map $p(t)$ and, provided that the average $\widehat p$ exists, consider the convergence functions defined for $\ep>0$, 
\[
\delta(\ep)=\sup_{\substack{ 0\le \ep R\le 1}}\ep\,\bigg|\int_{0}^{R}\big(p(t)-\widehat p\,\big)\, dt\bigg| \quad \hbox{and}\quad 
\widetilde\delta(\ep)=\sup_{\substack{t_0\geq 0 \\[1.5pt] 0\le \ep R\le 1}}\!\!\ep\,\bigg|\int_{t_0}^{t_0+R}\big(p(t)-\widehat p\,\big)\, dt\bigg|\,.
\]
The first case to be considered is that of bounded integrals, that is, we assume that $p$ satisfies  $\sup_{R\geq 0} \left| \int_0^R \big( p(t)-\widehat p\,\big)\,dt\right|<\infty$  for some $\widehat p\in\R$. In this case,  $p(t)$ is UKBM with average $\widehat p$ and  $\widetilde\delta(\ep)=O(\ep)$ as $\ep\to 0$, using again the standard Landau's notation. This is the case for all periodic maps and for some almost periodic maps, but let us mention that in fact the property of bounded integrals is not frequent in almost periodic maps, from a precise topological point of view. 

After the previous easy but very restrictive case, we introduce arguments of topological dynamics and ergodic theory in order to better understand the existence of averages and the behaviour of the convergence functions. Let us assume that $p(t)$ is also uniformly continuous and let us build its hull $H(p)=\mathrm{cls}\,\{p_\tau\mid \tau\in\R\}$, as in \eqref{eq:hull} in Section \ref{subsec:attr-repel-pair}, where the closure is taken for the compact-open topology and $p_\tau:\R\to\R$ is the time-translation   $t\in\R\mapsto p_\tau(t)=p(t+\tau)$. Clearly, every element in $H(p)$ is also a bounded and uniformly continuous function. 
Recall that $H(p)$ is a compact metric space and the time-shift map 
$\sigma:\R\times H(p)\to H(p)$, $(t,h)\mapsto h_t$ defines a continuous flow. The flow $(H(p),\sigma)$ is \emph{minimal} if for every $h\in H(p)$, the trajectory $\{h_t\mid t\in\R\}$ is dense in $H(p)$. 

The Krylov–Bogoliubov theorem guarantees the existence of a normalised invariant measure for the compact flow on $H(p)$. 
Since the existence of the average only takes into account the values of the map $p(t)$ for $t\geq 0$, from a dynamical perspective it is natural to relate it with the omega-limit set of $p$ inside the hull 
\[
\omega(p):=\Big\{h\in H(p)\ \Big| \ h=\lim_{n\to\infty}p_{t_n},\text{ for some } (t_n)_{n\geq 1}\to \infty\Big\}\,.
\]
The set $\omega(p)$ is compact and invariant. Then, also for the flow $(\omega(p),\sigma)$  the set $\M_{\rm{inv}}(\omega(p),\sigma)$ of normalised invariant measures on $\omega(p)$ is nonempty. If this set is a singleton, we say that $\omega(p)$ is uniquely ergodic, since the unique invariant measure is then ergodic. 
It is well-known that, if $p$ is almost periodic, then $(H(p),\sigma)=(\omega(p),\sigma)$ is minimal and uniquely ergodic.


In order to apply Birkhoff's ergodic theorem, we consider the  evaluation function, 
\begin{equation}\label{eq:eval}
    E:H(p)\to \R\,,\quad h\mapsto h(0)\,,
\end{equation}
which is continuous and allows to recover the values that each element $h\in H(p)$ attains along its trajectory through the flow $(H(p),\sigma)$. Indeed, $E(h_t)=h_t(0)=h(t)$, $t\in\R$, for all $h\in H(p)$. In particular, $E(p_t)=p(t)$, $t\in\R$ recovers the values of the initial map $p(t)$. 

 If $\omega(p)$ is uniquely ergodic,  then the map $p$ is UKBM.  Let $\mu$ be the unique ergodic invariant measure. Then, a consequence of Birkhoff's ergodic theorem  (see \cite{cornfeld1982ergodictheory}) implies that there exist the following limits, uniformly for $t_0\geq 0$,  
\[
\lim_{R\to\infty} \frac{1}{R}\int_{t_0}^{t_0+R}  E(p_t)\,dt =\lim_{R\to\infty} \frac{1}{R}\int_{t_0}^{t_0+R} p(t)\,dt=  \int_{\omega(p)}E \,d\mu\,.
\]
Note that this is the case if $p$ is almost periodic.  Some examples of maps within this situation are bounded and uniformly continuous maps that are asymptotically constant, so that the omega-limit set $\omega(p)$ is a singleton and the unique ergodic measure is a Dirac measure. However, in these cases $p\notin \omega(p)$, the approximation to the omega-limit set can be very slow and the convergence function can exhibit different rates. For instance, the maps defined for $t\geq 0$ by 
\[
p_1(t)=\frac{t}{1+t^2}\,,\qquad    p_2(t)=\frac{ 1}{(1+t)^\alpha} \;\, (0<\alpha<1)\,, 
\]
have null average with $\delta_1(\ep)\sim  \ep |\log(\ep)|$  and $\delta_2(\ep) = \ep^\alpha/(1-\alpha)$ as $\ep\to 0$, respectively (see \cite{sanders2007averaging}). Also for almost periodic maps $p(t)$ the convergence functions can be not as good as $O(\ep)$. There are examples in the literature of almost periodic functions $p(t)$ with null average whose integrals $\int_0^R p(t)\,dt\geq c\,R^{1-\alpha}$ as $R\to\infty$ for $0<\alpha<1$ (see \cite{zhikovLevitan1977,ortega2006}, among others). In these cases, $\delta(\ep) \geq c\,\ep^{\alpha}$ as $\ep\to 0$.

Finally, if $\omega(p)$ is not uniquely ergodic, then Birkhoff's ergodic theorem cannot be applied as above. 
It might still happen that for the evaluation function there is a constant $c\in\R$ such that
\[
\int_{\omega(p)}E\,d\mu=c \quad\text{for all } \,\mu\in \M_{\rm{inv}}(\omega(p),\sigma)\,,
\]
implying that the previous conclusions remain valid and $p$ is UKBM. For example, this can occur for maps with bounded integrals and several ergodic measures supported in $\omega(p)$. However, if $(\omega(p),\sigma)$ is not uniquely ergodic, 
there are typically two ergodic measures $\mu_1,\mu_2\in\M_{\rm{inv}}(\omega(p),\sigma)$ such that $\int_{\omega(p)}E\,d\mu_1\neq\int_{\omega(p)}E\,d\mu_2$. Since every map in $\omega(p)$ is the limit of a sequence of maps along the semitrajectory $\{p_t\mid t\ge0\}$  by construction, there are sequences of real numbers $(t^1_n)_{n\geq 1},(R^1_n)_{n\geq 1},(t^2_n)_{n\geq 1},(R^2_n)_{n\geq 1}$, with $t^i_n,R^i_n\to\infty$ as $n\to\infty$ for $i=1,2$, such that
\[
\bigg|\frac{1}{R^1_n}\int_{t^1_n}^{t^1_n+R^1_n}p(t)\,dt-\int_{\omega(p)}E\,d\mu_1\bigg|<\frac{1}{n}\]
and\[
\bigg|\frac{1}{R^2_n}\int_{t^2_n}^{t^2_n+R^2_n}p(t)\,dt-\int_{\omega(p)}E\,d\mu_2\bigg|<\frac{1}{n}
\]
for every $n\geq 1$. From Birkhoff's ergodic theorem, it follows the existence of an invariant subset  with complete measure $H_r\subset H(p)$ of regular points (see Mañé \cite{mane1987ergodic}), i.e.,~for every $h\in H_r$ there is an ergodic measure $\mu_h\in\M_{\rm{inv}}(\omega(p),\sigma)$ such that the following limit exists
\[
\lim_{R\to\infty} \frac{1}{R}\int_{0}^{R} E(h_t)\,dt=\lim_{R\to\infty} \frac{1}{R}\int_{0}^{R}h(t) \,dt=\int_{\omega(p)}E\,d\mu_h\,.
\]
That is, for every $h\in H_r$,   the function $h$ is KBM. In particular, if $p\in H_r$, then $p$ itself is KBM, although not UKBM. Additionally, note that for a time-average to approximate sufficiently well each $\mu_i$, 
increasingly bigger values $t^i_n$ and $R^i_n$ are expected.
Then the convergence function $\delta(\ep)$ goes to zero in a slower fashion as $\ep\to 0$. If, on the other hand, $p\notin H_r$, the existence of the limit \eqref{eq:average-0} could still hold---in which case the previous conclusions would hold too---but it is in general not to be expected. 

In conclusion, we can affirm that UKBM functions are the natural environment for performing an averaging method.

\subsection{Averaging  applied to the EBM close to the upper hyperbolic solution}\label{subsec:averaging-hyperbolic}

In this section, we profit from the dynamical structure of the EBM and its  averaged problem, detailed in Section \ref{subsec:attr-repel-pair}, in order to analyse the error caused by averaging. The presence of locally attracting solutions with exponential rate of asymptotic convergence for each problem allows us to extend the approximation due to averaging on a compact interval to the whole positive half-line, in invariant regions around the hyperbolic solutions.
Results of this type are originally due to Sanchez–Palencia~\cite{sanchez1975methode,sanchez1976methode} and Eckhaus~\cite{eckhaus1975new}. We hereby adapt the proof of \cite[Theorem 5.5.1]{sanders2007averaging} to the EBM. However, we note that the averaged model is not able to reproduce the qualitative behaviour of the solutions of the nonautonomous EBM that is linked to the temporal variation of the coefficients. The arguments used in this section are necessarily technical in order to provide a rigorous proof of our statements.

\begin{figure}[h]
    \centering
    \begin{overpic}[width=\textwidth]{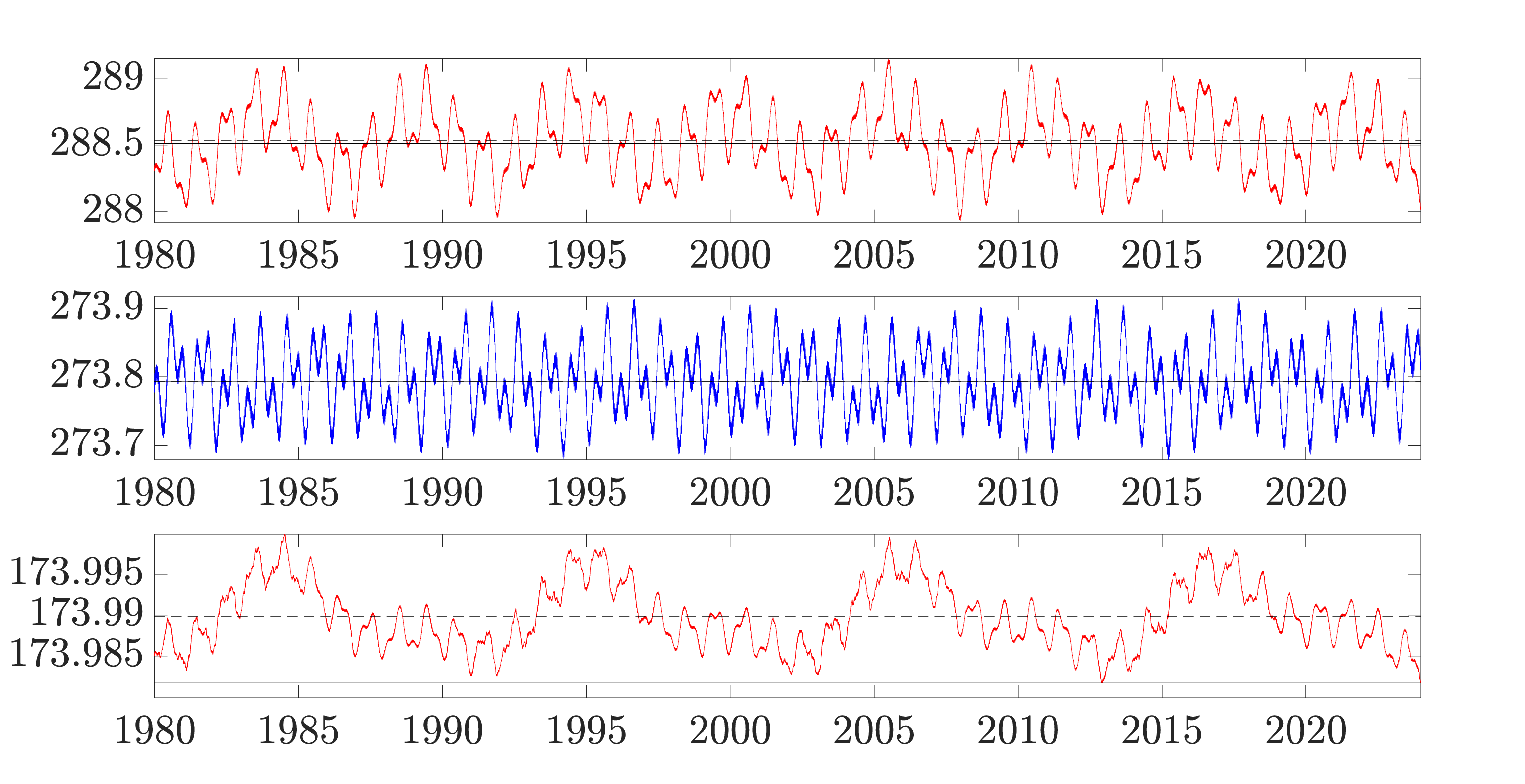}
     \put(95,25){\begin{turn}{90} $T\, (K)$ \end{turn}}
    \put(43,-1){$t$ (years)}
    \end{overpic}
    \caption{The hyperbolic solutions  in \eqref{eq: atractor a_1} of the quasi-periodic EBM: $a_1(t)$ above, $r(t)$ in the middle and $a_2(t)$ below. The maps $a_1(t)$ and $a_2(t)$ are depicted in red solid lines (attractors) and $r(t)$ in a blue solid line (repeller). Their averages between 1900 and 2024 are depicted in dashed lines. The hyperbolic equilibria of the averaged model are depicted in  black solid lines. Note that the three panels do not have the same scale on the vertical axis.}
    \label{fig:sols+avg}
\end{figure}

Assume that $f(t,T)$ is KBM and consider the EBM $T'=\ep_0 f(t,T)$ in \eqref{eq:Ghil-years-ep} and the averaged problem $T'=\ep_0 \widehat f(T)$ in  \eqref{eq:averaged EBM}. 
As explained in Section \ref{subsec:averaged-EBM}, we can assume that $f$ is bounded, Lipschitz in the variable $T$ uniformly for $t\in\R$, and $f(t,T)=0$ if $T\leq 0$ or $T\geq 500$ for all $t\in\R$.  Then, thanks to the compact support, we can just consider the gauges $\delta(\ep)$ and $\widetilde\delta(\ep)$ given in \eqref{eq:convergence-function-delta} and \eqref{eq:convergence-function-delta-hat}, respectively, where $\sup_{T}$ is taken over all $T\in\R$. 

For $\ep>0$, let $T_\ep(t,0,T_0)$ and $T_\ep^*(t,0,T_0)$ be, respectively, the solutions of the  Cauchy problems \eqref{eq:23/01-16:45} and \eqref{eq:23/01-16:46}, which we intend to compare, in particular for the value $\ep=\varepsilon_0$. The theory of averaging provides estimates of the error $\sup_{t\in [0,1/\ep]}|T_\ep(t,0,T_0)-T_\ep^*(t,0,T_0)|$ on long time intervals $[0,1/\ep]$. Generally, the averaging error is of order $O\big(\sqrt{\delta(\ep)}\big)$ on the interval $[0,1/\ep]$ (see \cite[Theorem 4.3.6]{sanders2007averaging} for the precise statement), but  improved error estimates such as $O\big(\delta(\ep)\big)$ can be obtained under additional conditions (see \cite[Theorem 4.5.5]{sanders2007averaging}). 

Note that, although the results of averaging are typically written for initial conditions at $0$,  this is not restrictive since, if the initial condition is given at a different time $T(t_0)=T_0$, one can  look at the translated map $(T_\ep)_{t_0}(t)=T_\ep(t+t_0)$, which is a solution of the translated equation $T'=\ep\,f_{t_0}(t,T)$ with $(T_\ep)_{t_0}(0)=T_\ep(t_0)$. In fact, when $f(t,T)$ is UKBM, then the averaging theory provides estimates of the error $\sup_{t\in [t_0,t_0+1/\ep]}|T_\ep(t,t_0,T_0)- T_\ep^*(t,t_0,T_0)|$ uniformly for $t_0\geq 0$ in terms of the gauge $\widetilde\delta(\ep)$, just as before. 

Precisely, the results in this section apply to EBMs \eqref{eq:Ghil-years-ep} with bounded and uniformly continuous time varying coefficients $I(t)$ and $m(t)$ that are UKBM, and such that the map of semiequilibria and concavity zones in 
Figure~\ref{fig:semiequilibria} (left figure) applies. Basically, this happens if $I(t)$ and $m(t)$ vary between the same bounds as those of the quasi-periodic model in Section \ref{subsec:nonautonomous-forcing}; see also Figure \ref{fig:conc-conv}. 
Then, as explained in Section \ref{subsec:attr-repel-pair}, the EBM  and its averaged problem respectively admit an attractor-repeller pair within the interval of temperatures $[273.5\, K,305\, K]$, which lies in the upper region of concavity: see $a_1(t)$  in \eqref{eq: atractor a_1} and let us denote by $T^*_a$ the exponentially stable equilibrium 
of the autonomous model \eqref{eq:averaged EBM} inside $[275\, K,305\, K]$.  

Consider a temperature interval $I_0=[T_1,T_2]\subset [275,300]$ that contains both the equilibrium $T^*_a$ and the attracting solution $a_1(t)$, $t\in \R$ of the EBM, and determines a positively invariant band for both problems (in particular, this happens if $g_*(T_1)> 0$ and $g^*(T_2)< 0$ for the maps in Figure \ref{fig:conc-conv}). 
Let us fix a time $\tau_0\geq 0$ and let us introduce, for each $\ep>0$ and $h>0$, the formula of the averaging error on the interval $[t_0,t_0+h/\ep]$ uniformly for $t_0\geq \tau_0$ and for initial temperatures in $I_0$, 
\begin{equation}\label{eq:K ep unif}
\mathcal{E}_{I_0}(\ep,h):=\sup_{\substack{t_0\geq \tau_0,\, T_0\in I_0\\[3pt]t\in[t_0,t_0+h/\ep]}} |T_\ep(t,t_0,T_0)- T_\ep^*(t,t_0,T_0)|\,.
\end{equation}
The averaging theory says that $\mathcal{E}_{\{T_0\}}(\ep,h)\to 0$ as $\ep\to 0$, for each $T_0$ and $h>0$ fixed, and the rate of convergence can be expressed in terms of the gauge $\widetilde\delta(\ep)$, as it was indicated before. On the other hand, if now $\ep>0$ is fixed, $\mathcal{E}_{I_0}(\ep,h)\leq 2\,\|f\|_\infty h\to 0$ as $h\to 0$: just note that we can bound $|T_\ep(t,t_0,T_0)- T_\ep^*(t,t_0,T_0)|\leq 2\,\|f\|_\infty h$   by adding and subtracting $T_0$ and applying the Mean Value Theorem twice. 

Our main  interest is in the averaging error for the fixed value of $\ep=\ep_0$, when the solutions are those of the EBM and its averaged model. For this reason, from now on we shall simply denote by $T(t,t_0,T_0)$ and  $T^*(t,t_0,T_0)$ the solutions of \eqref{eq:Ghil-years-ep} and \eqref{eq:averaged EBM}, respectively. 
Also, for simplicity, we reuse the notation  $T'=g(t,T)$ for the EBM \eqref{eq:Ghil-years} in Section~\ref{secintro}, so that  the averaged EBM \eqref{eq:averaged EBM} is written as  $T'=\widehat g(T)$. 

By the monotonicity of the solutions of scalar ODEs, for all $T_0\in I_0=[T_1,T_2]$, $T^*(t,0,T_1)\leq T^*(t,0,T_0)\leq T^*(t,0,T_2)$ and, since $\widehat g \,'$ is nonincreasing in the concave zone $[275,300]$, $\widehat g \,' (T^*(t,0,T_0))\leq \widehat g \,' (T^*(t,0,T_1))$ for all $t\geq 0$. Now, because $\widehat g$ is of class $C^1$ and $\widehat g \,'(T^*_a)<0$,  by the robustness of the property of uniform asymptotic stability for linear equations, the linear variational equation along the solution $T^*(t,0,T_1)$, that is,  $T'=\widehat g \,' (T^*(t,0,T_1))\,T$, is also uniformly asymptotically stable (note that $T^*(t,0,T_1)\to T^*_a$ as $t\to\infty$). Therefore, there exists an $\ell_0>0$ (which implicitly  depends on $\ep_0$, and on the  band's endpoint $T_1$ of $I_0$) such that 
\begin{equation}\label{eq:lema}
    \exp \int_0^t \widehat g \,' \big(T^*(s,0,T_0)\big)\,ds \leq \frac{1}{2} \quad\hbox{for all }\,T_0\in [T_1,T_2]\;\,\hbox{and }\, t\geq \ell_0\,.
\end{equation}
Then, it follows that for all $t_0\geq \tau_0$ and  all initial conditions within the band $T_{0,1},T_{0,2}\in [T_1,T_2]$,
\begin{equation}\label{eq:contract-pairs}
    |T^*(t,t_0,T_{0,1})-T^*(t,t_0,T_{0,2})|\leq \frac{1}{2}\, |T_{0,1}-T_{0,2}| \quad \hbox{for all }\, t\geq t_0 +\ell_0\,.
\end{equation}
To see it, first note that the problem $T'=\widehat g(T)$ is autonomous and then use the Mean Value Theorem to write  
\begin{align*}
    |T^*&(t,t_0,T_{0,1})-T^*(t,t_0,T_{0,2})| = |T^*(t-t_0,0,T_{0,1})-T^*(t-t_0,0,T_{0,2})| \\ 
& = \bigg|\int_0^1 \frac{\partial T^*}{\partial T_0} \big(t-t_0,0, \lambda\,T_{0,1}+(1-\lambda)\,T_{0,2}\big)\,d\lambda  \bigg|\,|T_{0,1}-T_{0,2}|\\
& = \bigg|\int_0^1 \bigg(\exp \int_0^{t-t_0} \widehat g \,' \big(T^*(s,0, \lambda\,T_{0,1}+(1-\lambda)\,T_{0,2})\big)\,ds \bigg)\,d\lambda  \bigg|\,|T_{0,1}-T_{0,2}|\\
&\leq \frac{1}{2}\, |T_{0,1}-T_{0,2}|\,,
\end{align*}
where \eqref{eq:lema} has been applied in the last inequality for each $\lambda\in [0,1]$. We remark that we have chosen the value $1/2$ for the contraction process, but any value of $0<\gamma<1$ would work equally well.

For the value of $\ep_0$, taking $h_0 = \ep_0\,\ell_0>0$, we affirm that the uniform averaging error bound in \eqref{eq:K ep unif} on a positively invariant band of temperatures around the hyperbolic attractor,  can be extended to the whole half-line by doubling it. Namely, 
\begin{equation}\label{eq:whole half-line}
\sup_{\substack{t_0\geq \tau_0,\, T_0\in I_0\\[3pt]t\in[t_0,\infty)}} |T(t,t_0,T_0)- T^*(t,t_0,T_0)|\leq 2\,\mathcal{E}_{I_0}(\ep_0,h_0)\,.
\end{equation} 
In order to  prove this fact, we shall use a summation trick due to Sanchez-Palencia \cite{sanchez1975methode}, following the proof of \cite[Theorem 5.5.1]{sanders2007averaging}. For each $t_0\geq \tau_0$, consider the partition of the positive half-line 
\[
[t_0,\infty)=\bigcup_{j\geq 0}\Big[t_0+j\,\frac{h_0}{\ep_0},t_0+(j+1)\,\frac{h_0}{\ep_0}\Big]=:\bigcup_{j\geq 0}J_j\,.
\]
For $t\in J_0$ we have \eqref{eq:K ep unif}. Now, if we assume that 
\begin{equation}\label{eq:induction}
\sup_{t\in J_j }|T(t,t_0,T_0)-T^*(t,t_0,T_0)|\le \bigg(\sum_{n=0}^j \frac{1}{2^n}\bigg)\,\mathcal{E}_{I_0}(\ep_0,h_0)\,,
\end{equation}
then, for $t\in J_{j+1}$ we can write 
\begin{align*}
    |&T(t,t_0,T_0)-T^*(t,t_0,T_0)|\le \nonumber  \\ 
    &\quad      |T(t,t-\ell_0,T(t-\ell_0,t_0,T_0))-T^*(t,t-\ell_0,T(t-\ell_0,t_0,T_0))|\nonumber \\
    &\qquad+|T^*(t,t-\ell_0,T(t-\ell_0,t_0,T_0))-T^*(t,t-\ell_0,T^*(t-\ell_0,t_0,T_0))| \\ 
    &\quad\le \mathcal{E}_{I_0}(\ep_0,h_0) +\frac{1}{2}\,|T(t-\ell_0,t_0,T_0)-T^*(t-\ell_0,t_0,T_0)| \nonumber\\ 
    &\quad\le\,  \bigg(1 +\frac{1}{2}  \sum_{n=0}^j \frac{1}{2^n} \bigg)\,\mathcal{E}_{I_0}(\ep_0,h_0) = \bigg(\sum_{n=0}^{j+1} \frac{1}{2^n}\bigg)\,\mathcal{E}_{I_0}(\ep_0,h_0)\,,\nonumber
\end{align*}
where we have applied \eqref{eq:K ep unif} and \eqref{eq:contract-pairs} in the second-to-last inequality and the induction hypothesis \eqref{eq:induction} in the last one.  Therefore, we can conclude that the formula in \eqref{eq:induction}  works for all $j\geq 0$ and since $\sum_{n=0}^\infty \frac{1}{2^n}=2$, and this has been done for each $t_0\geq \tau_0$, the averaging error estimate in \eqref{eq:whole half-line} holds.

To finish this section, we get some estimations of $l_0$ and $h_0$ for the quasi-periodic EBM. Since the derivative  $\widehat g \,'$ is decreasing in the concave zone and the band $[T_1,T_2]$ is invariant, $\widehat g \,' (T^*(s,0,T_0))\leq \widehat g \,' (T_1)$ for all $T_0\in [T_1,T_2]$, $t\geq \ell_0$ and $s\in [0,t]$. Thus, as far as $\widehat g \,'(T_1)\leq 0$, we can just take $\ell_0$ such that $\exp (\widehat g \,' (T_1) \,\ell_0)\leq 1/2$ in order to fulfil condition \eqref{eq:lema}. Note that, according to Figure \ref{fig:conc-conv}, at some temperature $T_1^*$ in between $275$ and $276$, $\widehat g \,'(T_1^*)=0$ and there is a change of sign in the derivative. In fact, we can get close to this point, and still get reasonable values for $\ell_0$ keeping $h_0$  really small. 
We collect some values in Table \ref{table h y l}, taking invariant symmetric bands with respect to the averaged temperature $288.5$ of the attracting solution $a_1(t)$, containing both $a_1(t)$ and $T^+_a$---see Figure \ref{fig:sols+avg}. For the calculations we have just taken $\ep_0 = 1.2 \,{\cdot}\, 10^{-3}$. 

\begin{table}[H]
    \centering
\begin{tabular}{c c c c}
$[T_1,T_2]$ & $\widehat g \,'(T_1)$ & $\ell_0$ & $h_0$ \\ \hline
$[280, 297]$ & -0.8568 & 0.809 & 0.000971 \\
$[276,301]$ & -0.4620   & 1.500  & 0.001800 \\
$[275.7,301.3 ]$ & -0.1583   & 4.379 & 0.005254\\
\end{tabular}  
\caption{Some values of $\ell_0$ and $h_0$ for different invariant bands in the quasi-periodic EBM}
\label{table h y l}
\end{table}

\section{Climate models response and sensitivity}\label{sec response sensitivity}

The notions of climate response and sensitivity are markedly autonomous as they rely on the existence of an attracting equilibrium. Emerging from \cite{charney1979}, the equilibrium climate sensitivity (ECS, for short) has been defined as the increase in global mean surface temperature due to radiative forcing change after the fast-acting feedback processes in the earth system reach equilibrium (e.g., see \cite{ashwin2020extreme,bastiaansenashwinheydt2023climate, knutti2017beyond,rugenstein2021three}; also see \cite{chekrounetal2011stochasticclimate,dijdstraViebahn2015,senior2000time} for  different approaches).  However, equilibria, in the sense of stationary solutions, are rare objects for truly nonautonomous dynamical systems. It is therefore necessary to accordingly extend these notions. 

\subsection{A nonautonomous version of two-point response functions}\label{subsec:R1-R2}
We use the two-point climate response and sensitivity introduced by \cite{vonderheydtashwin2016,ashwin2020extreme}
as a basis and inspiration for our definitions. Let $T'=g(t,T)$ and $T'=\widetilde g(t,T)$ be two alternative  energy balance models for the average atmospheric temperature of the planet, which is considered to be {\it the observable} in our climate systems. For a given  initial state $T_0\in\R^+$ at initial time $t_0\in\R$, we define the two-point response function $R_1(\cdot,t_0,T_0):[t_0,\infty)\to \R^+$ as the map that compares the temperature evolution as time passes, according to the solutions $T(\cdot,t_0,T_0)$ and $\widetilde T(\cdot,t_0,T_0)$  of the respective equations: 
\begin{equation}\label{eq:def-R1}
R_1(t,t_0,T_0)=\widetilde T(t,t_0,T_0)-T(t,t_0,T_0)\,,\quad t\geq t_0\,.
\end{equation}
Note that, since constant solutions are not expected, both solutions are evaluated after the lapse of time $t-t_0$, to take into account the effect of the variation of the vector fields with time. 

Under standard regularity assumptions on the vector fields, the response function $R_1$ is continuously   differentiable, and for all $t_0\in\R$ and $T_0\ge 0$,
\begin{equation}\label{eq:R1-properties}
   R_1(t_0,t_0,T_0)=0\,,\quad\text{and}\quad R_1'(t_0^+\!,t_0,T_0)=\widetilde g(t_0,T_0)-g(t_0,T_0)\,,
\end{equation}
where $R_1'(t_0^+\!,t_0,T_0)$ denotes the right derivative of $R_1(t,t_0,T_0)$ at $t_0$. On the other hand, it is clear that, in general, the limit of $R_1(t,t_0,T_0)$ as $t\to\infty$ does not exist. Nonetheless, if we assume that both models fall within the dynamical description in Section \ref{subsec:attr-repel-pair},   the asymptotic behaviour of $R_1$ can be described in terms of the asymptotic behaviour of the global hyperbolic attractors $a_1(t)$ and $\widetilde a_1(t)$, $t\in\R$ respectively of $T'=g(t,T)$ and $T'=\widetilde g(t,T)$ in the band of temperatures $T_0\in[275,305]$. From \eqref{eq: atractor a_1} applied to both models we can deduce that there exist constants $c_0>0$ and $\alpha>0$ such that   
\begin{equation}\label{eq:R1-asymptotic}
|R_1(t,t_0,T_0)-(\widetilde a_1(t)-a_1(t))|\le c_0\,e^{-\alpha (t-t_0)},\quad \text{for }\,t\ge t_0\,,\; T_0\in [275,305]\,.
\end{equation}
In other words, for sufficiently large $t>t_0$ the function $R_1(t,t_0,T_0)$ approximates the dynamical behaviour of $\widetilde a_1(t)-a_1(t)$ independently of the initial condition $T_0\in [275,305]$.

In addition to  the response function $R_1$, it is reasonable to define an "average" two-point response function $R_2$, whose limit as $t\to\infty$  is expected to be finite at least in the uniquely ergodic case. Precisely, we define the average two-point response function  $R_2(\cdot,t_0,T_0):(t_0,\infty)\to \R^+$ as
\begin{equation}\label{eq:def-R2}
\begin{split}
R_2(t,t_0,T_0)&=\frac{1}{t-t_0}\int_{t_0}^{t}\big|\widetilde T(s,t_0,T_0)-T(s,t_0,T_0)\big|\,ds\\&=\frac{1}{t-t_0}\int_{t_0}^{t}\big|R_1(s,t_0,T_0)\big|\,ds\,.
\end{split}
\end{equation}
The function $R_2$ is also continuous and differentiable in $t$. Moreover, by the fundamental theorem of calculus,  $R_2$ can be extended with continuity to $[t_0,\infty)$, since 
$\lim_{t\to t_0^+}R_2(t,t_0,T_0)=  |R_1(t_0,t_0,T_0)|=0$ for all $t_0\in\R$ and $T_0\in  \R^+$,
and  it is also right differentiable at $t_0$. It suffices to apply  L'Hôpital's rule and \eqref{eq:R1-properties} to write 
\begin{align*}
    R_2'(t_0^+\!,t_0,T_0)&:=\lim_{t\to t_0^+}\frac{R_2(t,t_0,T_0)}{t-t_0}= \lim_{t\to t_0^+}\frac{\displaystyle\int_{t_0}^{t}\big|R_1(s,t_0,T_0)\big|\,ds}{(t-t_0)^2}\\ &= \lim_{t\to t_0^+}\frac{|R_1(t,t_0,T_0)\big|}{2\,(t-t_0)}=\frac{1}{2}\,\big|R_1'(t_0^+\!,t_0,T_0)\big|=\frac{\big|\widetilde g(t_0,T_0)-g(t_0,T_0)\big|}{2}\,,
\end{align*}
for all $t_0\in\R$ and $T_0\in  \R^+$.
Additionally, provided that \eqref{eq:R1-asymptotic} holds, by the definition of $R_2$ in \eqref{eq:def-R2} we can bound
\[
\begin{split}
\bigg| &R_2(t,t_0,T_0) -  \frac{1}{t-t_0}\int_{t_0}^{t}\big|\widetilde a_1(s)-a_1(s) \big|\,ds \bigg| \\ 
&\le   \frac{1}{t-t_0}\int_{t_0}^{t} \left| \big|R_1(s,t_0,T_0)\big| - \big|\widetilde a_1(s)-a_1(s)\big| \right| ds\\
&\le \frac{1}{t-t_0} \int_{t_0}^{t}  \big|R_1(s,t_0,T_0)-\big(\widetilde a_1(s)-a_1(s)\big) \big|\, ds    \le  c_0\,\frac{1-e^{-\alpha (t-t_0)}}{\alpha\,(t-t_0)}\,.
\end{split}
\]
Consequently, if the map $\big|\widetilde a_1(t)-a_1(t) \big|$ is UKBM, then the limit as $t\to\infty$ of $R_2(t,t_0,T_0)$ exists and its value is independent of $t_0$ and $T_0\in [275,305]$. This issue has been addressed in Section \ref{subsec:dynamical underst}. If the omega-limit set of this map for the shift flow is uniquely ergodic, the map is UKBM. This happens, for instance, if both attracting maps $a_1(t)$ and $\widetilde a_1(t)$ are quasi-periodic, as it turns out in quasi-periodic models. In this case, if the omega-limit set is an $n$-dimensional torus and $\mu$ is the unique ergodic measure, then, for the evaluation map defined in \eqref{eq:eval}, due to Birkhoff`s ergodic theorem it holds that
\begin{equation*}
\lim_{t\to\infty} R_2(t,t_0,T_0)=\int_{\mathbb{T}^n} E \,d\mu\,.
\end{equation*}
If the omega-limit set $\omega(p)$ of the map $p(t)=\big|\widetilde a_1(t)-a_1(t) \big|$ is not uniquely ergodic but still $p(t)$ is KBM, then for each $(t_0,T_0)$,   $\lim_{t\to\infty} R_2(t,t_0,T_0)$ exists and lies in the finite interval $[r_1,r_2]$ for 
\[
r_1 := \inf_{\mu} \int_{\omega(p)} E \,d\mu \quad \hbox{ and }\quad r_2:=\sup_{\mu} \int_{\omega(p)} E \,d\mu\,,
\]
where both the $\inf$ and the $\sup$ are taken over all the normalised invariant measures in  $\omega(p)$.

\begin{figure}
    \centering
    \begin{overpic}[width=\textwidth]{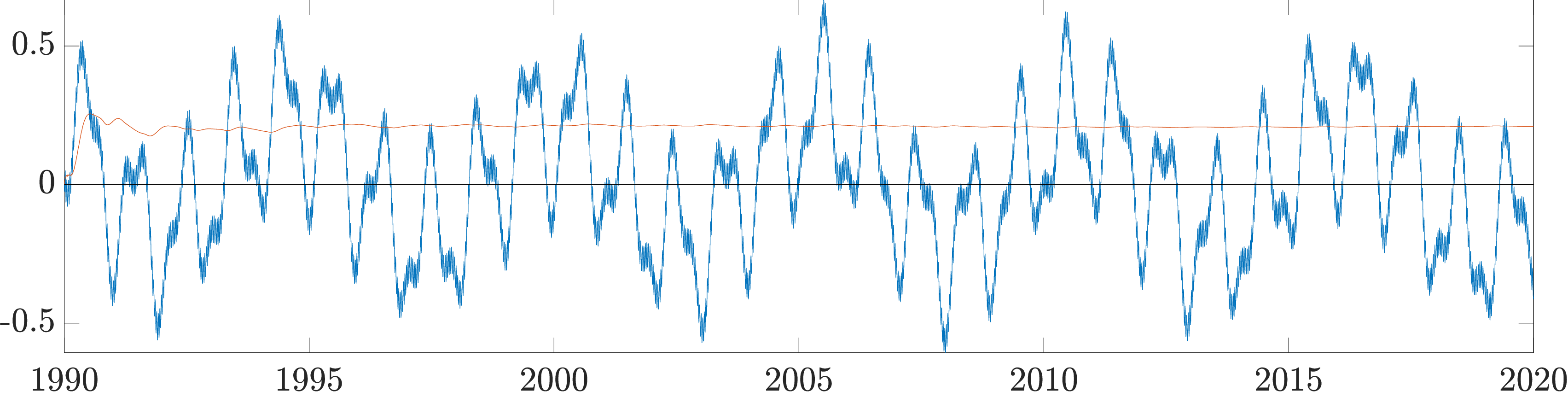}
    \end{overpic}
    \caption{$R_1(t,t_0,T_0)$ (in blue) and $R_2(t,t_0,T_0)$ (in orange) for the quasi-periodic and average models starting at $t_0=1990$ with $T_0$ the attracting equilibrium of the averaged problem and $t\in[1990,2020]$.}
    \label{fig:sols+avg 2}
\end{figure}

In particular, one can use the above defined response functions to compare the dynamics of the nonautonomous EBM $T'=g(t,T)$ in \eqref{eq:Ghil-years}, assuming that $g(t,T)$ is a KBM function, and its averaged counterpart $T'=\widehat g(T)$ introduced in Section \ref{subsec:averaged-EBM}. 
Note that in this case $a_1(t)$ is the hyperbolic solution representing the current state of the climate whereas $\widetilde a_1(t)\equiv T^*_a$ is the hyperbolic equilibrium of the autonomous averaged problem. 
Then, \eqref{eq:R1-asymptotic} implies that $R_1(t,t_0,T_0)$ asymptotically behaves as $T^*_a-a_1(t)$ and therefore it shows the same recurrency properties of $g$. The asymptotic behaviour of $R_2(t,t_0,T_0)$ is instead related to the behaviour of the average of $|T^*_a-a_1(t)|$. Numerical evidence of such asymptotic behaviours for the model with the quasi-periodic forcing presented in Section \ref{subsec:nonautonomous-forcing} is shown in Figure~\ref{fig:sols+avg 2}.

\subsection{Modelling CO$_2$ forcing in the nonautonomous EBM}

A typical assumption is that the forcing due to emissions acts additively with respect to \eqref{eq:Ghil-years}. In other words, we consider the differential equation
\begin{equation}\label{eq:EBMwF}
   T'= g(t,T) + F (C(t)) \,,
\end{equation}
where
\begin{equation}\label{eq:RadiativeForcingFromCarbon}
\begin{split}
     F(C) = a_0\left(a_1 -a_2(C - C_0)^2 + a_3(C - C_0) -a_4\sqrt{N}\right) \log(C/C_0)\,
\end{split}
\end{equation}
with $a_0=1.05$, $a_1=5.2488$, $a_2=2.48\cdot10^{-7}$, $a_3=7.59\cdot10^{-4}$, $a_4=2.15\cdot10^{-3}$, $C$ represents equivalent concentration of CO$_2$ ($CO_2\hbox{-eq}$) in ppm (parts per million) and $C_0$ is a baseline equivalent concentration of CO$_2$ in ppm so that $F(C_0)=0$~\cite{montzka2023noaa}.
Since our model \eqref{eq:Ghil-years} is adjusted to have its attracting solution around the average surface temperature of 288.5 K in $2022$, the formula \eqref{eq:RadiativeForcingFromCarbon}  is hereby considered with $C_0=530$ ppm---the value of $CO_2\hbox{-eq}$ in 2022,  extracted from \cite{montzka2023noaa}. 
For the sake of simplicity, we shall neglect the contribution of $N$, which represents the abundance of $N_2O$ in ppb (parts per billion).

\begin{figure}[h]
\hfill
\begin{overpic}[width=0.96\textwidth]{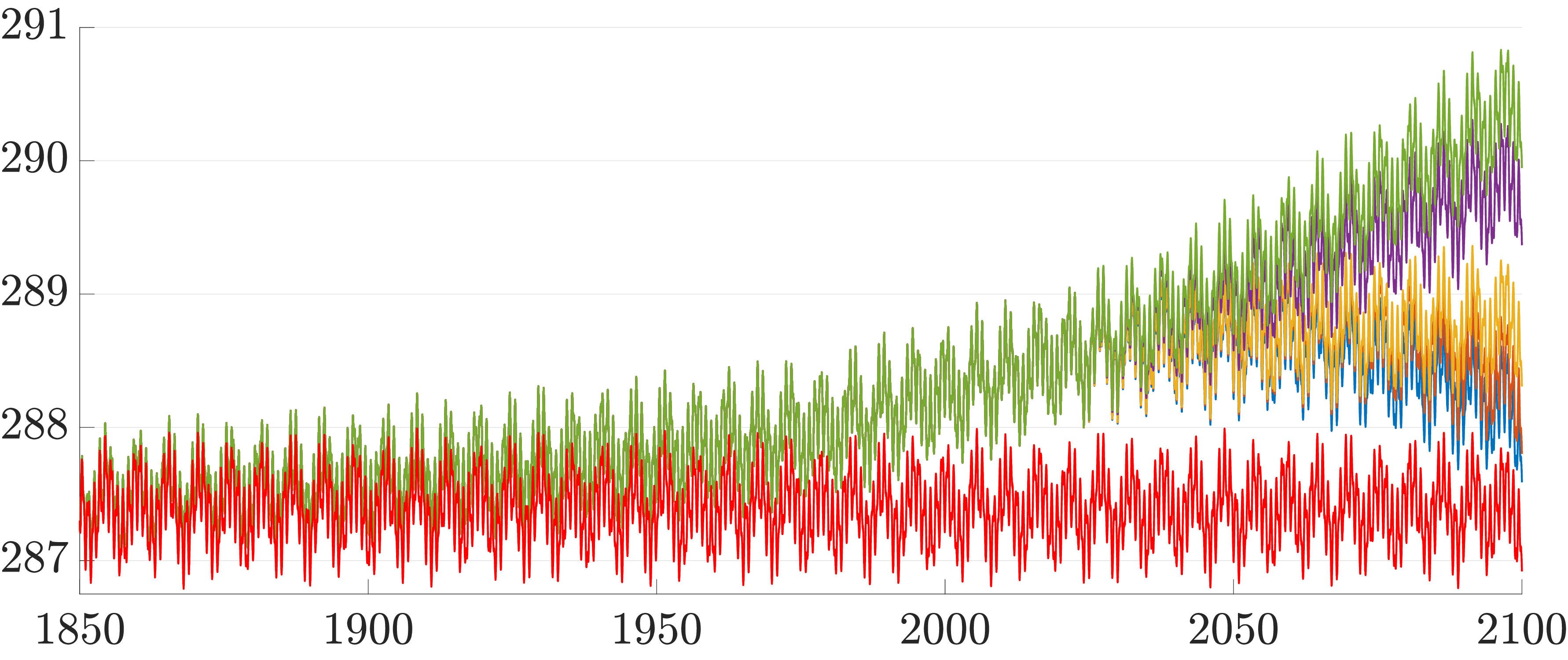} 
\put(45,-2){$t$ (years)}
\put(-3,20){\begin{turn}{90} $T\, (K)$ \end{turn}}
\end{overpic}
\caption{The attracting solutions of the EBM with CO$_2$ forcing are superimposed and compared to the attracting solution of \eqref{eq:EBMwF} with $C\equiv 278.3$ ppm---preindustrial CO$_2$-eq concentration---in red. Blue corresponds to SSP1--1.9, orange to SSP1--2.6, yellow to SSP2--4.5, purple to SSP3--7.0 and green to SSP5--8.5.}
\label{fig:SSPs-temperature}
\end{figure}

In order to assess the physical consistency of \eqref{eq:EBMwF}, we note that \eqref{eq:RadiativeForcingFromCarbon} assumes a negative value $F(278.3)=-3.3441$ at the preindustrial level of carbon dioxide concentration of 278.3 ppm---value also extracted from \cite{montzka2023noaa}. Arguing as in Step 1 in Section \ref{subsec:attr-repel-pair}, the model \eqref{eq:EBMwF} with constant term $F(278.3)$ has an attractor-repeller pair in the band of temperatures $[273.5,305]$. Just note that the value of $g_*(275.5)$ is close to $5.8127$, the relative maximum value of $g_*$ (see Figure \ref{fig:conc-conv}), so that a map of semiequilibria and concavity regions similar to the one in Figure \ref{fig:semiequilibria} (left figure)  applies to the present induced skew-product flow. 
Moreover, the relevant attractor of \eqref{eq:EBMwF} with constant term $F(278.3)$ has numerical average of 287.4 K, which is roughly $1.1$ K less than the average of the attractor of \eqref{eq:Ghil-years} (see also Fig.~\ref{fig:SSPs-temperature}). Therefore, \eqref{eq:EBMwF} and \eqref{eq:RadiativeForcingFromCarbon} together appear to provide a reasonable approximation of the expected dynamics in the years 1850 and 2022.

We shall consider two types of time-dependent profiles for the carbon dioxide emissions  $C(t)$ in \eqref{eq:EBMwF}-\eqref{eq:RadiativeForcingFromCarbon}. 
The first type dates back to Charney \cite{charney1979} and consists of performing an instantaneous doubling (or quadrupling, etc.) of the preindustrial concentration of CO$_2$ in the atmosphere,
\[
C(t)=\begin{cases}
    278.3 & \text{if } t\le 1850\\
    2^\gamma\cdot 278.3 & \text{if } t> 1850
\end{cases}\,, \qquad\hbox{for}\;\, \gamma=1,2,3\,.
\]
From a mathematical standpoint, the instantaneous doubling (or quadrupling, etc.) gives rise to a discontinuity in time for the differential equation \eqref{eq:EBMwF}. Although this discontinuity is fundamentally innocuous, it does require a bit of care when constructing the skew-product flow, in that the differential equation cannot be treated as a standard one but it needs to be understood in the context of Carathéodory differential equations \cite{longo2018topologies}. Roughly speaking, this means that when reading \eqref{eq:EBMwF} we need to think of the equivalent integral problem
\[
T(t)=T(t_0)+\int_{t_0}^t\big[g(s,T(s))+F(C(s))\big]\,ds\,
\]
and choose a suitable integral topology for the construction of the hull \cite{longo2018topologies,longo2019weak}.

The second type of time-dependent profile for $C(t)$  simulates five emissions scenarios of the possible future evolution of the climate on Earth depending on the emission of anthropogenic drivers of climate change, mitigation strategies and their impact on the human population.  The five scenarios are called Shared Socioeconomic Representative Pathways (SSPs) and described in the 6th assessment report (AR6) on climate change of the Intergovernmental Panel for Climate Change (IPCC)~\cite{ipcc2023climate}. Their qualitative features are summarised in Table~\ref{table:SSPs}.

The high and very high greenhouse gases emissions scenarios (SSP3-7.0 and
SSP5-8.5) have CO$_2$ emissions that roughly double from current levels by 2100 and 2075, respectively. The
intermediate greenhouse gases emissions scenario (SSP2-4.5) has CO$_2$ emissions remaining around current levels until the
middle of the century. The very low and low greenhouse gases emissions scenarios (SSP1-1.9 and SSP1-2.6) have CO$_2$
emissions declining to net zero around 2050 and 2070, respectively, followed by varying levels of net negative CO$_2$ emissions.

The pathways are labelled as SSPx-y, where 'SSPx' refers to the Shared Socioeconomic Pathway describing the socioeconomic trends underlying the scenarios, and ‘y’ refers to the level of radiative forcing (in watts per square metre, or $W\cdot m^{-2}$) resulting from the scenario in the year 2100.

\begin{table}
    \centering
    \begin{tabular}{ |p{1.45cm}||p{4cm}|p{4.1cm}|p{2cm}|  }
 \hline
 \textbf{Name}& \textbf{Radiative Forcing} &\textbf{Concentration}&\textbf{Pathway shape}\\[1ex]
 \hline
  SSP1--1.9  & ${\approx}\,1.9\; Wm^{-2}$
 in 2100 with limited or no overshooting   &${\approx}\,400$ CO$_2$-eq in
2100 &   Peak around 2050 and decline 
\\[4ex]
SSP1--2.6& ${\approx}\,2.6 \;Wm^{-2}$
 in 2100 with high overshooting  &${\approx}\,450 $  CO$_2$-eq in
2100&   Peak around 2070 and decline 
\\[4ex]
 SSP2--4.5&  ${\approx}\,4.5 \;Wm^{-2}$
 at stabilization
after 2100  & ${\approx}\,650 $  CO$_2$-eq
(attained at stabilization after 2100)   &Stabilization; no
overshoot 
\\[4ex]
 SSP3--7.0& ${\approx}\,7 \;Wm^{-2}$
 in 2100 & ${\approx}\,1035 $ CO$_2$-eq
in 2100 & Rising 
\\[4ex]
 SSP5--8.5    & ${\approx}\,8.5\; Wm^{-2}$ 
 in 2100  & ${\approx}\,1370 $ CO$_2$-eq in 2100  &  Rising\\[1ex]
 
 \hline
\end{tabular}
\vspace{1mm}
    \caption{Socio-Economic Representative Pathways SSP1 -- SSP5 equivalent radiative forcing, CO$_2$-eq concentration and pathway shape. The approximate radiative forcing levels were defined as $\pm5\%$ of the stated level in $Wm^{-2}$. Radiative forcing values include
the net effect of all anthropogenic greenhouse gases and other forcing agents. The CO$_2$-eq concentrations were calculated with the simple
formula concentration = $277.15 \,{\cdot}\, \exp(\hbox{forcing}/5.2488)$ on the base of \eqref{eq:RadiativeForcingFromCarbon} and the analogous calculation in~\cite[page xiii]{moss2008towards}. 
}
\label{table:SSPs}
\end{table}

The five scenarios are implemented by considering a shape-preserving piecewise cubic interpolation of the assumed average CO$_2$-eq concentration before the industrial revolution, estimated in 278.3 ppm, the available yearly data from 1979 till 2023 in \cite{montzka2023noaa}, and the qualitative information on the future of the SSPs reported in Table \ref{table:SSPs}. 
Note that the cumulative radiative forcing does not exactly match the ones of the SSPs at the year 2100 because $N$ in \eqref{eq:RadiativeForcingFromCarbon} has been neglected.

Furthermore, also the EBM \eqref{eq:EBMwF} with time-dependent forcing $F(C(t))$ has an attracting solution in the band of temperatures of interest, for any of the pathways of evolution mentioned above for carbon dioxide emissions (see Figure \ref{fig:SSPs-temperature}), including abrupt ones. The arguments to prove it are similar to those used in Step 2 in  Section~\ref{subsec:attr-repel-pair}. Let us explain it with a bit of detail. First, note that the forcing term does not affect the zones of concavity of the initial EBM $T'=g(t,T)$ in \eqref{eq:Ghil-years} but the hull of $g(t,T)+F(C(t))$ is different from the hull of $g(t,T)$ built in Section \ref{subsec:attr-repel-pair}. We assume that $C(t)$ remains constant before the preindustrial time and after the year 2100 for the hull construction, and we use a weak topology (see \cite{longo2024critical}) for the jump functions modelling an abrupt increase of CO$_2$, since in this case the equation is in the context of Carathéodory~\cite{longo2018topologies,longo2019weak}, as already mentioned before. The map $F$ is nondecreasing for the values of C we are considering. Thus, $F(C(t))$ has its minimum value at $F(278.3)=-3.3441$ and its maximum value at, say, $F(C_1)$, and for all the maps $h(t,T)$ in the new hull it holds that $0<g_*(275.5)-3.341\leq h(t,275.5)$ and $h(t,T_1)\leq g^*(T_1)+F(C_1) <0$ for all $t\in\R$, for a sufficiently high temperature $T_1$. This implies that $[275.5, T_1]$ determines a positively invariant band for the induced skew-product flow.  Then, by cutting the vector field at the temperature $275.5$ and extending it below using a second order Taylor expansion at $275.5$, we get a strictly concave and coercive vector field that does have an attractor-repeller pair, the attractor lying in the zone of interest and the repeller below being meaningless in this context. Figure \ref{fig:SSPs-temperature} shows the attracting solutions of the EBM with the different  CO$_2$ forcing terms corresponding to the five SSPs superimposed and compared to the attracting solution of \eqref{eq:EBMwF} with $C\equiv 278.3$ ppm---preindustrial CO$_2$-eq concentration.

\subsection{Response functions and climate sensitivity in EBMs with CO$_2$ forcing}\label{sec:response-and-sensitivity}
In this section, the response functions $R_1$  \eqref{eq:def-R1} and $R_2$ \eqref{eq:def-R2} are used to compare the EBM $T'= g(t,T)+F (278.3)$ for the map $g$ given in \eqref{eq:Ghil-years} with the  perturbed model $T'= g(t,T) +F (C(t))$, when a time-dependent forcing $F(C(t))$ intervenes. 

We note that our approach is coherent with the response to forcing, and equilibrium (or asymptotic) response for autonomous models, as introduced in \cite{bastiaansenashwinheydt2023climate}. In fact, if an autonomous model $T'=g(T)$ has a hyperbolic attracting  equilibrium at $T_a$ and the model is perturbed by a (possibly time-dependent) forcing $\Delta f(t)$, that is, $T'=g(T)+\Delta f(t)$ with solutions  $\widetilde T(t,t_0,T_0)$, $t\geq t_0$, then, by taking $T_0=T_a$, the response function $R_1(t,t_0,T_a)=\widetilde T(t,t_0,T_a)-T_a$ suits the definition in \cite{bastiaansenashwinheydt2023climate}, as well as its asymptotic behaviour given by $\lim_{t\to\infty} R_1(t,t_0,T_a)$ (note that the limit may not be well-defined) suits the notion there called equilibrium response. 
In fact, the asymptotics of the  response function $R_1$ reduces to the equilibrium climate sensitivity, ECS, considered in \cite{charney1979} when $g$ does not depend on time, $\Delta f$ is an abrupt doubling of the preindustrial concentration of CO$_2$ and $T_0$ is the stable equilibrium corresponding to the warm climate state in the preindustrial era. This is also the classical setup for most studies on climate sensitivity, which has become increasingly detailed and quantitative assessing a likely range of $1.5 - 4.5 \;K$, see \cite{knutti2017beyond}.  More recently, the study of ECS has also focused on the transient dynamics after abrupt doubling of CO$_2$ (see for example \cite{dai2020improved,ghil2020physics}).

\begin{figure}[h]
\centering
\begin{overpic}[width=0.47\textwidth]{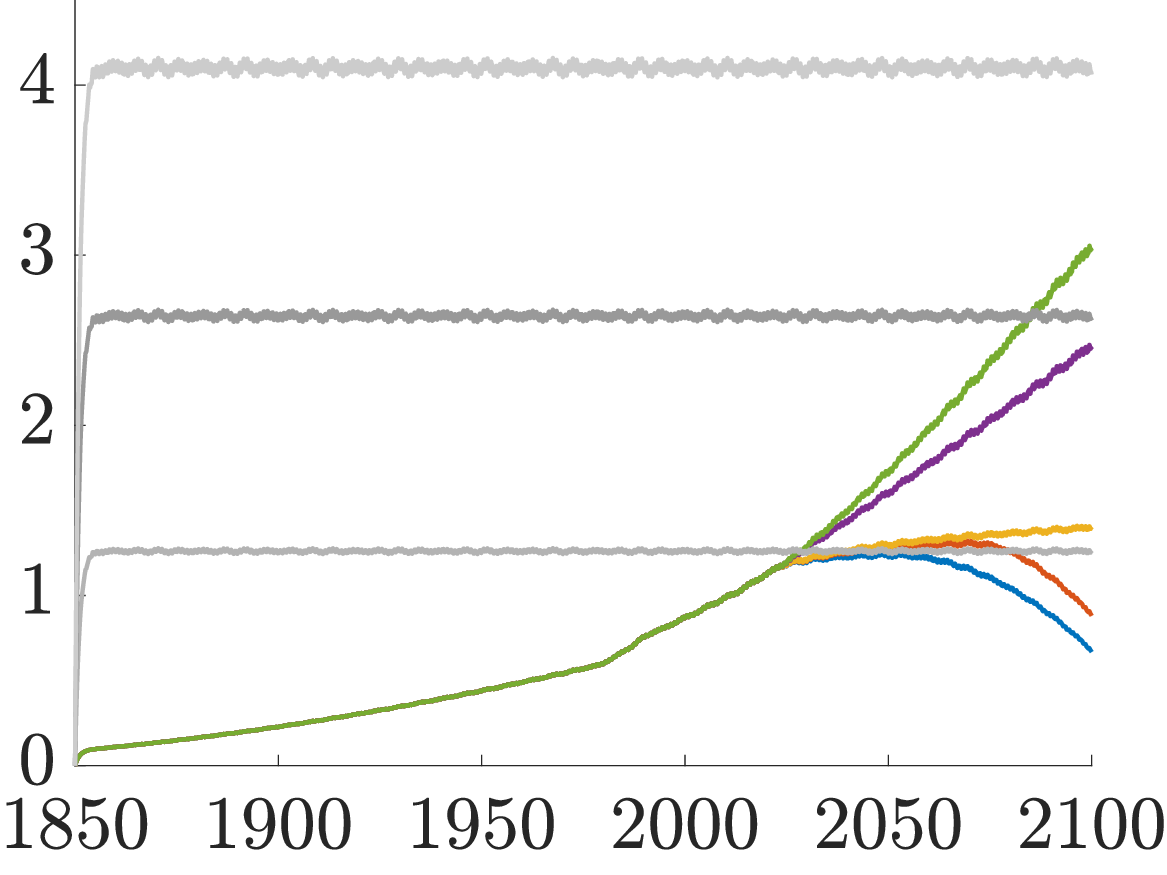} 
\put(45,-4){$t$ (years)}
\put(-6,40){$R_1$}
\put(10,30){$\gamma=1$}
\put(10,50){$\gamma=2$}
\put(10,72){$\gamma=3$}
\end{overpic}
\hspace{0.1cm}
\begin{overpic}[width=0.47\textwidth]{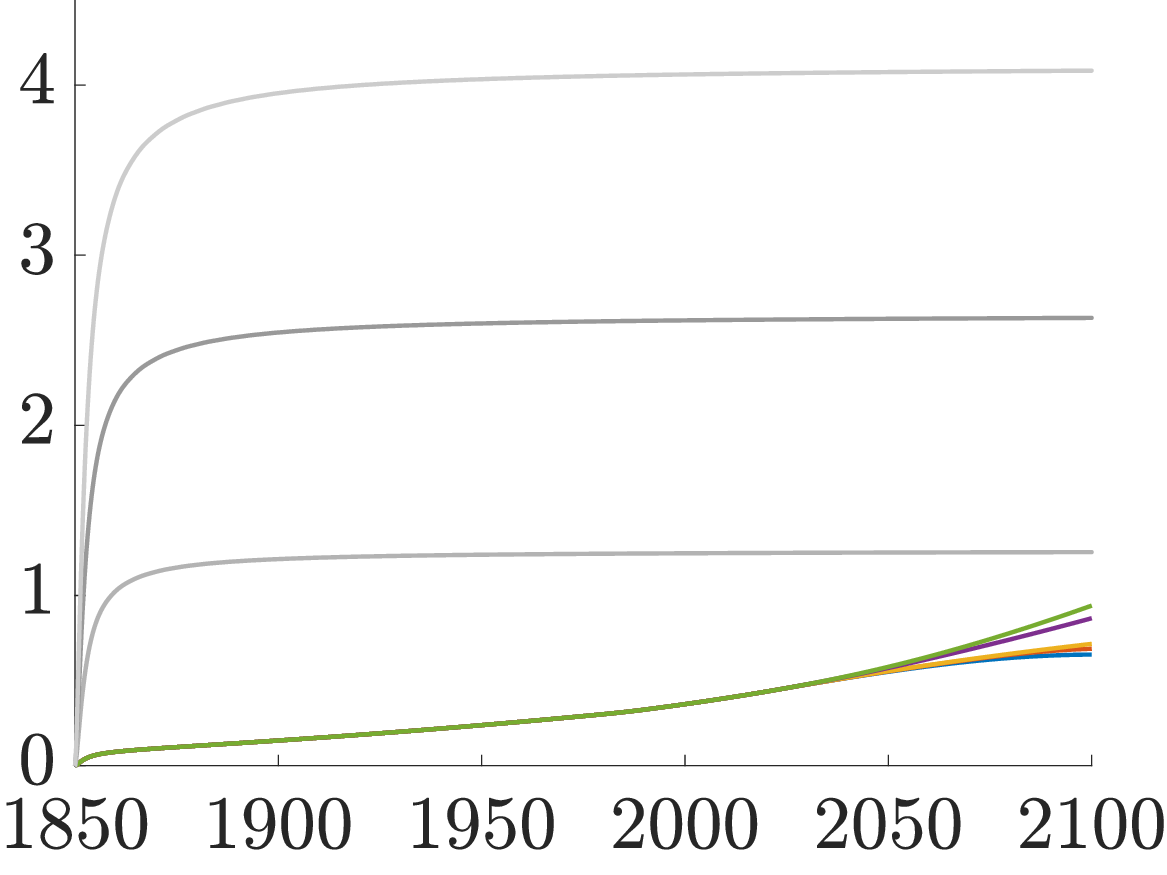} 
\put(45,-4){$t$ (years)}
\put(-6,40){$R_2$}
\put(80,30){$\gamma=1$}
\put(80,50){$\gamma=2$}
\put(80,72){$\gamma=3$}
\end{overpic}

\caption{Response functions $R_1$ on the left and $R_2$ on the right. Colors correspond to the SSPs. Grey curves to abrupt CO$_2$ increase by a factor $2^\gamma$ with $\gamma=1,2,3$.}
\label{fig:SSPs-response}
\end{figure}

Figure \ref{fig:SSPs-response} shows the graph of the response functions $R_1(t,t_0,T_0)$ (left-hand side) and $R_2(t,t_0,T_0)$ (right-hand side) as functions of time $t$, where $t_0=1850$ and $T_0$ is the numerically approximated value of the attractor of \eqref{eq:EBMwF} with $C=278.3$ in the year 1850 after a transient of 6850 years has been disregarded. The two-point response to the SSPs is shown maintaining the same color scheme of Figure \ref{fig:SSPs-temperature} and juxtaposed to three additional scenarios accounting for the abrupt increase of CO$_2$ by a factor $2^\gamma$, with $\gamma=1,2,3$ (depicted in shades of gray). 
The left panel is especially interesting, since it can be observed that the quasi-periodic oscillation almost completely cancels out in the difference of the temperatures as defined in \eqref{eq:def-R1}. This is an expected behaviour for differential equations as the magnitude of a perturbation tends to zero (see for example how oscillations on the fast timescale tend to disappear for the green curve close to the year 1850) due to continuous variation of solutions on compact intervals, but it does not generally extend to greater perturbations. 
There are, nonetheless, reasonable physical arguments supporting this model behaviour: in all the scenarios considered, the quasi-periodic forcing is identical. As a result, we expect the oscillations in the temperature profiles on the fast timescale to follow the same patterns of upward and downward movement, with only the amplitude varying. This amplitude increase becomes more pronounced as nonlinear effects begin to dominate. 

\begin{figure}[h]
\centering
\begin{overpic}[width=0.47\textwidth]{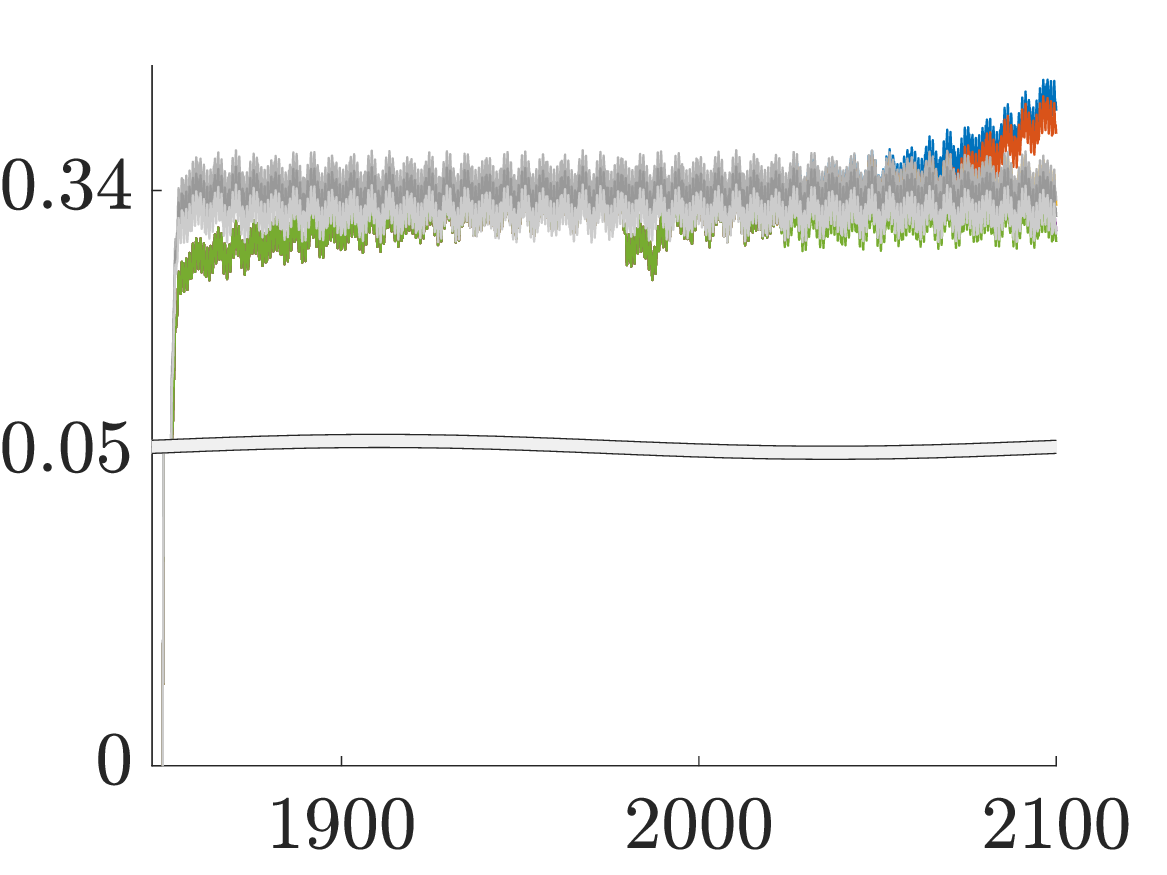} 
\put(45,-3){$t$ (years)}
\put(-6,40){$S_1$}
\end{overpic}
\hspace{0.1cm}
\begin{overpic}[width=0.47\textwidth]{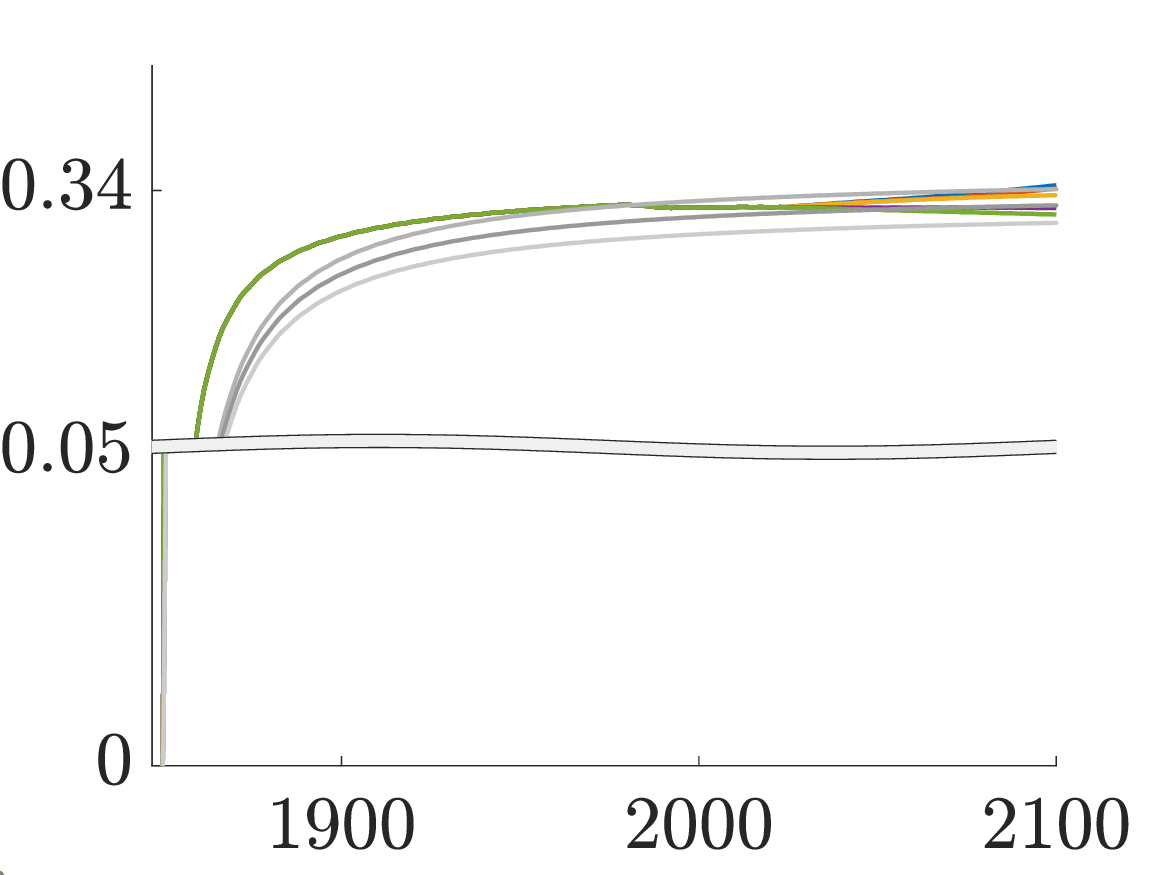} 
\put(45,-3){$t$ (years)}
\put(-7,40){$S_2$}
\end{overpic}

\caption{Sensitivity functions $S_1$ on the left and $S_2$ on the right. Colors correspond to the SSPs. Grey curves to abrupt CO$_2$ increase by a factor $2^\gamma$ with $\gamma=1,2,3$. The overlapping of all graphs shows a general agreement of the sensitivity functions despite the different types of forcings.}
\label{fig:SSPs-sensitivity}
\end{figure}

In order to study the variation of the response with respect to the forcing, we consider the notion of the two-point climate sensitivity parameter (see \cite{palaeosens2012making,ashwin2020extreme}),
\[
S_1(t,t_0,T_0):=\frac{R_1(t,t_0,T_0)}{F(C(t))-F(C(t_0))}\,,\quad t>t_0\,,
\]
which is well-defined as long as $\Delta F(C(t)):=F(C(t))-F(C(t_0))\neq0$ for $t>t_0$. Note that $S_1(t,t_0,T_0)$ can be extended to $t_0$ with continuity in all the cases under consideration: for the abrupt increase of CO$_2$, $\Delta F(C(t))$ is a positive constant for $t>t_0$, so that $S_1(t_0,t_0,T_0):=\lim_{t\to t_0^+} S_1(t,t_0,T_0)=0$. For the five emission scenarios in Table \ref{table:SSPs}, using L'Hôpital's rule and \eqref{eq:R1-properties} we have
\[
S_1(t_0,t_0,T_0):=\lim_{t\to t_0^+} S_1(t,t_0,T_0)=\lim_{t\to t_0^+}\frac{R_1'(t,t_0,T_0)}{F'(C(t))\,C'(t)}=\frac{0}{F'(C(t_0))\,C'(t_0)}=0
\]
because the last denominator is different from zero.
The numerical experiments for the five emission scenarios  and the abrupt CO$_2$ increase by a factor $2^\gamma$ with $\gamma=1,2,3$ are shown in the left-hand panel of Figure \ref{fig:SSPs-sensitivity}. It is possible to appreciate that the response in temperature roughly corresponds to 0.34 times the variation of forcing, highlighting a substantial proportionality for the considered range of values of CO$_2$. In fact, applying this relation on the time interval $[1900,2050]$, where the {\it linearity} is more evident, $R_1(t,t_0,T_0)\simeq 0,34 \,(F(C(t))-F(C(t_0)))$, one can deduce the same behaviour when any two of the perturbed models $T'=g(t,T)+F(C_1(t))$ and $T'=g(t,T)+F(C_2(t))$ are compared through the corresponding response function $\widetilde R_1(t,t_0,T_0)$, that is, $\widetilde R_1(t,t_0,T_0)\simeq 0,34 \,(F(C_1(t))-F(C_2(t)))$.  

An averaged two-point climate sensitivity parameter can be analogously defined using $R_2$ and the average variation of forcing,
\[\displaystyle
S_2(t,t_0,T_0):=\frac{R_2(t,t_0,T_0)}{\displaystyle \frac{1}{t-t_0}\int_{t_0}^{t}\Delta F(C(s))\,ds}\,,\quad t>t_0\,.
\]
Once more distinguishing the cases of abrupt CO$_2$ increase and the five emission scenarios, one can check that  $S_2$ can be extended with continuity to $t_0$ by taking $S_2(t_0,t_0,T_0)=0$. 
Numerical simulations for the five emission scenarios in Table \ref{table:SSPs} and the abrupt CO$_2$ increase by a factor $2^\gamma$ with $\gamma=1,2,3$ are shown in the right-hand panel of Figure \ref{fig:SSPs-sensitivity}.

\section{Random coefficients}\label{secrandom}
In this section, we clarify the construction of the chaotic coefficient used in Section \ref{subsec:nonautonomous-forcing} to emulate the behaviour of the cloud cover. As we shall rigorously explain, the starting point is in the chaotic features of the uniformly expanding circle maps. From this discrete process, we construct a continuous time flow via linear interpolation, which inherits the properties of chaos.

Let $\mathbb{S}^1=\R/(2\pi\Z)$ be the unit circle and consider the map $\varphi:\mathbb{S}^1\to\mathbb{S}^1$, $\theta\mapsto 2\theta \!\!\mod 2\pi$. A discrete forward dynamical system on $\mathbb{S}^1$ is generated by the iteration of $\varphi$. Note that backward extensions of semiorbits are nonunique for this semiflow. 
In particular, $(\mathbb{S}^1,\varphi)$ is chaotic in the sense of Devaney \cite[Example 8.6]{devaney2018introduction}. 
This means that the system satisfies the properties of sensitive dependence on initial conditions, density of periodic orbits  and topological transitivity. Since $\mathbb{S}^1$ is compact, topological transitivity is equivalent to  the existence of a positive semitrajectory that is dense (see 
Auslander and Yorke \cite{auslander1980interval}).  
Moreover, from \cite[Proposition I.11.4]{mane1987ergodic} 
it follows that the set $S\subset \mathbb{S}^1$ of initial data with dense semitrajectory is residual and, in particular, dense.

Let us consider the set $(\mathbb{S}^1)^\Z$ of bi-directional sequences of elements in $\mathbb{S}^1$. 
$(\mathbb{S}^1)^\Z$ endowed with the product topology is a compact metric space with distance between $\Theta=(\theta_n)_{n\in\Z}\in (\mathbb{S}^1)^\Z$ and $\widehat \Theta=(\widehat \theta_n)_{n\in\Z}\in (\mathbb{S}^1)^\Z$ defined by
\[
d(\Theta,\widehat\Theta)=\big|e^{i\theta_0}-e^{i\widehat \theta_0}\big|+\sum_{n=1}^{\infty}\frac{1}{2^{n}}\left(\big|e^{i\theta_n}-e^{i\widehat \theta_n}\big|+\big|e^{i\theta_{-n}}-e^{i\widehat \theta_{-n}}\big|\right),
\]
where $|\cdot|$ denotes the usual Euclidean distance in the plane. 
Next, we shall consider the closed subsets $C_+,C_-\subset(\mathbb{S}^1)^\Z$ made up of complete (respectively forward and backward) trajectories in $\mathbb{S}^1$ through $\varphi$, that is, the sets 
\[\begin{split}
    C_+&=\big\{(\theta_n)_{n\in\Z}\mid \theta_{n+1}=\varphi(\theta_n),\text{ for all } n\in\Z\big\}\,,\\
C_-&=\big\{(\theta_n)_{n\in\Z}\mid \theta_{n-1}=\varphi(\theta_n),\text{ for all } n\in\Z\big\}\,,
\end{split}
\]
with the induced topology. Note that one can define two homeomorphisms (i.e.,~continuous bijections with continuous inverse), the left shift $\Lambda_l:C_+\to C_+$ and the right shift $\Lambda_r:C_-\to C_-$ (see for example \cite{devaney2018introduction}) as the maps
\[
\Lambda_l\big((\theta_n)_{n\in\Z}\big)= (\theta_{n+1})_{n\in\Z}\,\quad\text{and}\quad
\Lambda_r\big((\theta_n)_{n\in\Z}\big)= (\theta_{n-1})_{n\in\Z}\,.
\]
It is easy to check that the map $J:C_+\to C_-$, $(\theta_n)_{n\in\Z}\mapsto (\theta_{-n})_{n\in\Z}$ is also a homeomorphism. Moreover, it is clear that
\[
J\circ\Lambda_l=\Lambda_r\circ J\,,
\]
where $\circ$ is the composition operator. Thus, the compact flows $(C_+,\Lambda_l)$ and $(C_-,\Lambda_r)$ are topologically conjugated.
Furthermore, $C_-$ is topologically conjugated to  the {\it inverse limit space} for the doubling map  $\varphi$, as originally introduced by R.F. Williams (see \cite[Section 2.5]{devaney2018introduction}). In turn, this inverse limit space is topologically conjugated to the solenoid in $\R^3$, which is a geometric model in the class of attractors known as expanding attractors, and is chaotic in the sense of Devaney. Thus, we obtain the same chaotic properties as well as the same geometric interpretation of $(C_+,\Lambda_l)$.

We shall now proceed to extend the dynamics of the uniformly expanding circle map $\varphi$ to a continuous real flow via linear interpolation. 
For every fixed $\Theta=(\theta_n)_{n\in\Z}\in C_+$ consider the complex function $p_\Theta:\R\to\mathbb{C}$ defined by 
\[
p_\Theta(t)= (t-\lfloor t\rfloor)e^{i\theta_{\lfloor t\rfloor +1}}+(1-t+\lfloor t\rfloor)e^{i\theta_{\lfloor t\rfloor}}, 
\]
where $\lfloor t\rfloor$ is the floor function of $t$, i.e.,~the greatest integer below $t$---note that $t-\lfloor t\rfloor\in [0,1)$. Additionally, consider the sets of complex functions
\[
\mathcal{H}_0=\big\{p_\Theta\mid\Theta\in C_+\big\}\quad\text{and}\quad \mathcal{H}=\big\{p_t\mid p\in \mathcal{H}_0\,,\, t\in[0,1]\big\}\,,
\]
where $p_t(s)=p(t+s)$ for every $s\in\R$, as usual. Endow both sets with the compact-open topology, which is also generated by the distance
\[
d(p,\widehat p)=\sup_{t\in[-1,1]}|p(t)-\widehat p(t)|+\sum_{n=1}^{\infty}\frac{1}{2^{n}}\sup_{t\in[n,n+1]\cup [-n-1,-n]}|p(t)-\widehat p(t)|\,.
\]
Now notice that the time-$1$ translation $\sigma_1:\mathcal{H}_0\to \mathcal{H}_0$, $p\mapsto \sigma_1(p)=p_1$ is continuous and it is topologically conjugated to the discrete semiflow $(C_+,\Lambda_l)$ via the homeomorphism 
\[
\widehat J:C_+\to \mathcal{H}_0,\quad \Theta\mapsto \widehat J(\Theta)=p_\Theta\,,\quad \text{which verifies}\quad \widehat J\circ \Lambda_l=\sigma_1\circ \widehat J\,.
\]
Let us now take into consideration the continuous-time shift on $\mathcal{H}$, $\sigma:\R \times \mathcal{H}\to \mathcal{H}$, $(t,p)\mapsto p_t$. It is easy to see that this map is well-defined and continuous. Thus, $(\mathcal{H},\sigma)$ is a continuous flow on a compact metric space. By construction, it inherits the property of chaos in the sense of Devaney.

\begin{figure}
\centering
\begin{overpic}[width=0.96\textwidth]{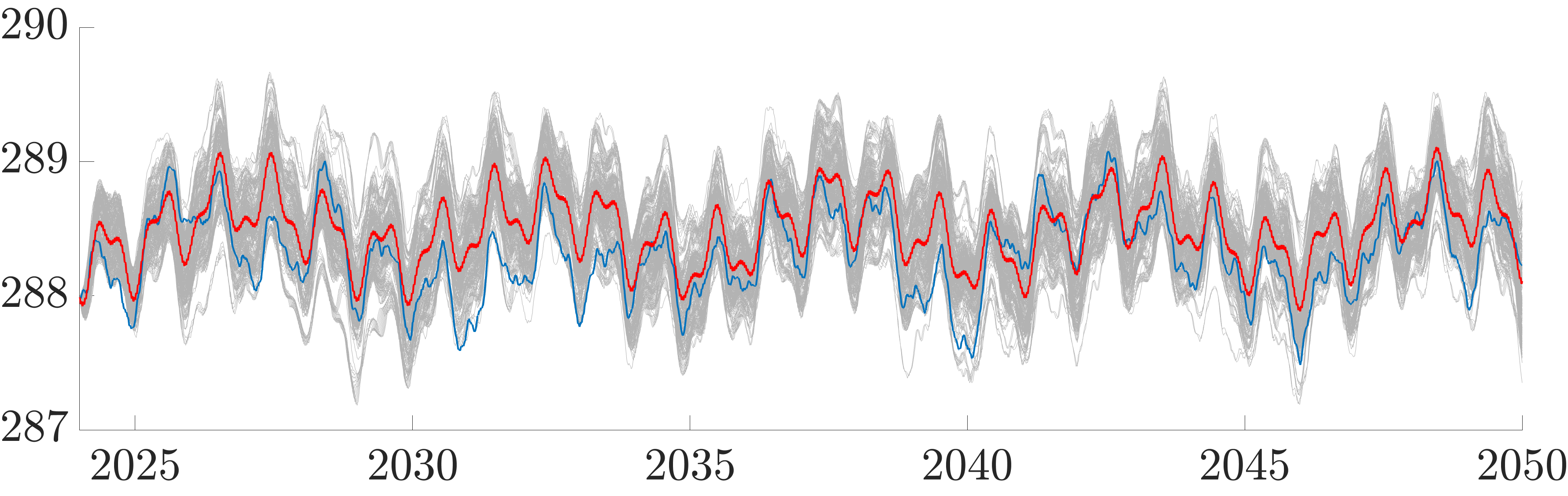} 
\put(45,-2){$t$ (years)}
\put(-3,13){\begin{turn}{90} $T\, (K)$ \end{turn}}
\end{overpic}

\caption{Graph of two hundred temperature profiles (in gray) solving the initial value problem \eqref{eq:Ghil-years}, with $T(2024)=T_0$ being the value of the hyperbolic attractor (in red) of the quasi-periodically forced differential equation at the year 2024, for different initial conditions on $\mathbb{S}^1=\R/(2\pi\Z)$ for the chaotic map $\varphi:\mathbb{S}^1\to\mathbb{S}^1$, $\theta\mapsto 2\theta \!\!\mod 2\pi$. Highlighted in blue is the initial condition $\theta_0=\pi\sqrt 5/2 \!\!\mod 2\pi$ for $\varphi$, which was also used to create the chaotic cloud profile in Figure \ref{fig:clouds}.}
\label{fig:chaotic}
\end{figure}

\begin{figure}
\centering
\begin{overpic}[width=0.96\textwidth]{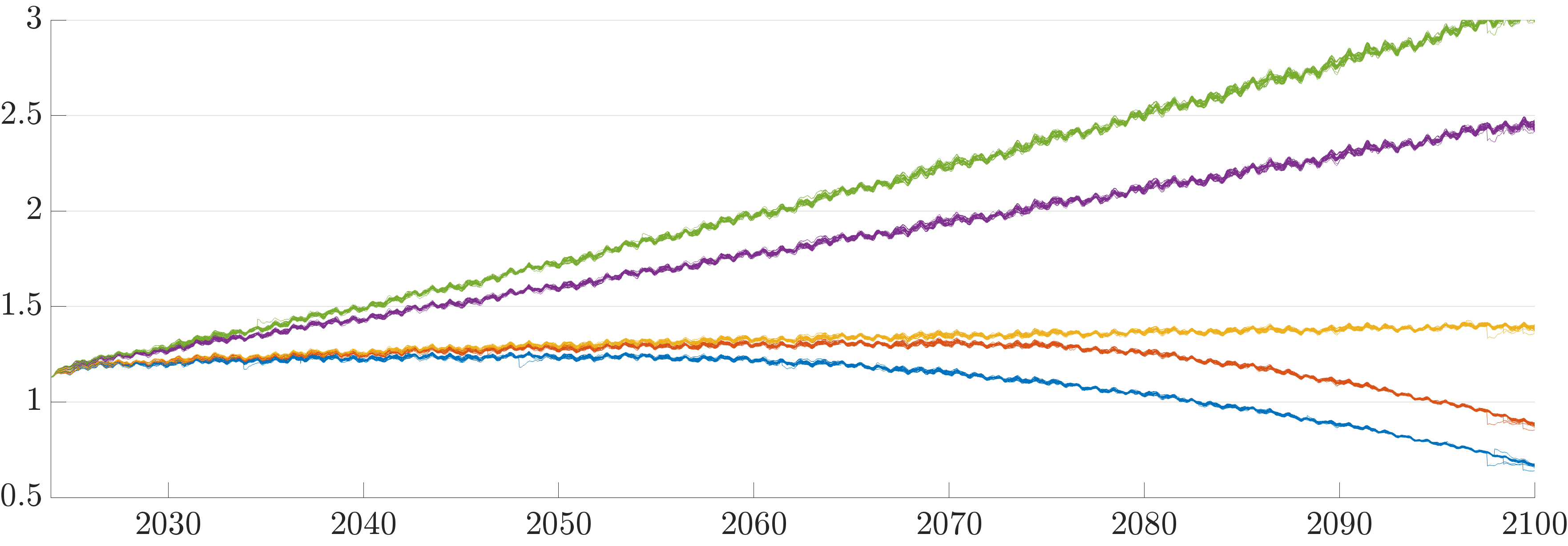} 
\put(45,-2){$t$ (years)}
\put(-3,20){$R_1$}
\end{overpic}\\
\begin{overpic}[width=0.975\textwidth]{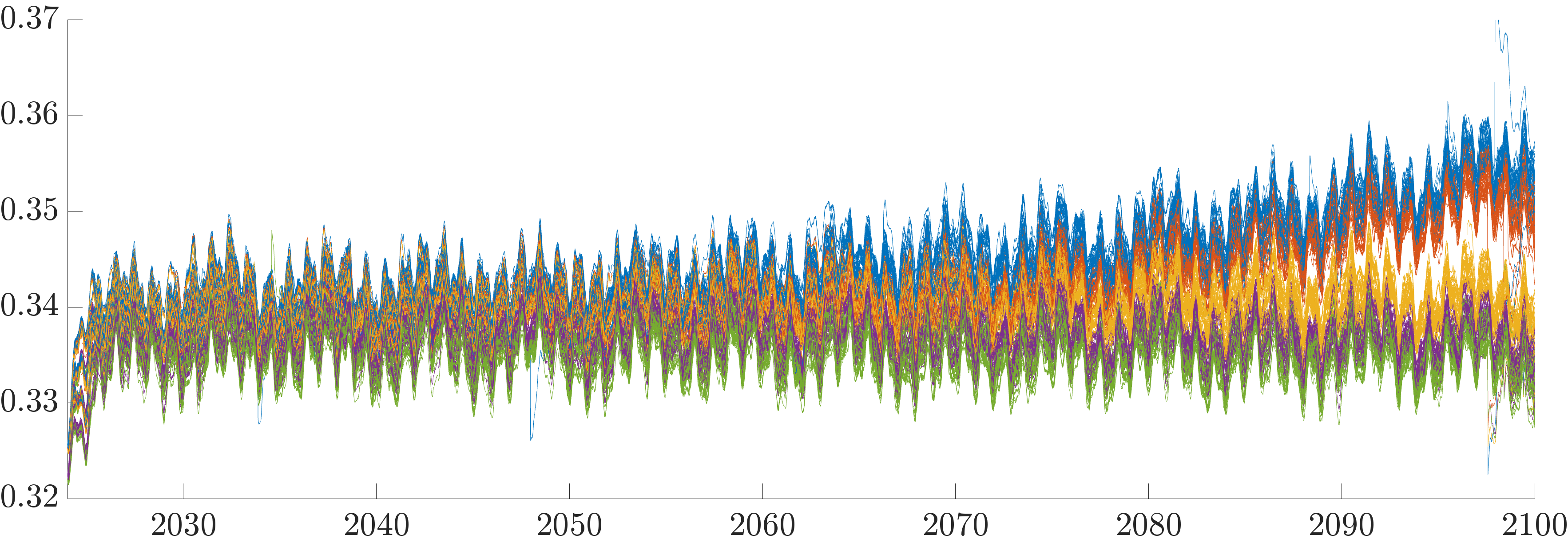} 
\put(45,-2){$t$ (years)}
\put(-3,20){$S_1$}
\end{overpic}

\caption{The upper panel shows the response functions with respect to the preindustrial  model $T'= g(t,T) + F (278.3)$  for fifty temperature profiles for each SSP in table \ref{table:SSPs} (Each color identifies one specific SSP).
The fifty temperature profiles for each SSP correspond to fifty uniformly distributed initial conditions on $\mathbb{S}^1=\R/(2\pi\Z)$ for the chaotic map $\varphi:\mathbb{S}^1\to\mathbb{S}^1$, $\theta\mapsto 2\theta \!\!\mod 2\pi$ forcing in \eqref{eq:Ghil-years}. The lower panel shows the parameter sensitivity function for the same setup. Note that the two panels have different scales on the vertical axis from the ones in Figures \ref{fig:SSPs-response} and \ref{fig:SSPs-sensitivity}, respectively. }
\label{fig:chaotic 3}
\end{figure}

The reasoning above allows to conclude the existence of a residual set $\mathcal{G}\subset\mathcal{H}$ such that for every $p\in\mathcal{G}$ the semitrajectory $\{p_t\mid t\ge 0\}$ is dense in $\mathcal{H}$. This means that such semitrajectory (randomly) approximates every other function in $\mathcal{H}$ on intervals of any arbitrary length. More precisely, chosen $\widehat p\in\mathcal{H}$ and a sequence of positive real numbers $(T_n)_{n\geq 1}$, with $T_n\to\infty$, there is a sequence of positive times $(t_n)_{n\geq 1}$, with $t_n\to\infty$, such that
\[
\sup_{t\in[0,T_n]} |p(t_n+t)-\widehat p(t)|<\frac{1}{n}\quad \text{for all } n\geq 1\,.
\]
It is obvious that such function $p$ cannot be UKBM as it does not admit an average and therefore, the methods of averaging are not applicable for this type of function. This information is particularly relevant in the context of climate, where randomness and chaoticity are essential ingredients of the dynamics. Therefore, a rigorous autonomous approximation of an EBM cannot be attained (in general) via averaging methods. However, not everything is lost: in Figure \ref{fig:chaotic} the upper attracting hyperbolic solution of the quasi-periodically forced EBM (representing current climate) is depicted in red and compared with the corresponding solutions of two hundred chaotically perturbed instances of the EBM (in gray) obtained using different initial conditions for the continuous extension of the uniformly expanding map presented above. The solution of one of such chaotically forced EBM is highlighted in blue. It is possible to appreciate that such solution combines a time-dependent behaviour which is remnant of the quasi-periodic forcing, with a more erratic behaviour due to the chaotic forcing. The gray cloud of chaotic trajectories closely surrounds the quasi-periodic one, providing numerical evidence of the robustness of hyperbolic solutions: given an $\ep>0$ there exists a $\delta=\delta(\ep)>0$ such that, if the equation with a hyperbolic solution is perturbed by a term of norm up to $\delta$, then the perturbed equation also has a hyperbolic solution which is located within distance $\ep$ of the initial hyperbolic solution. 
In the context of climate dynamics, this means that it is worth singling out the main frequencies to construct a quasi-periodic EBM (to which the methods of averaging can be applied) and then exploit hyperbolicity to extend the error estimates to the whole real line (Section \ref{subsec:averaging-hyperbolic}) and to ``small'' random forcing.
In Figure \ref{fig:chaotic 3} we also show the response and sensitivity functions for the forced EBM \eqref{eq:EBMwF} varying the chaotic term in the cloud cover.  \smallskip
 
To finish, let us remark that the chaotic flow $\mathcal{H}$ built upon the uniformly expanding circle map $\varphi$ can be embedded into a much larger set $\mathcal{P}$, which is also chaotic in the sense of Devaney (see Dueñas et al.~\cite[Theorem 3.8]{duenas2025Saddle}). Namely, for appropriate constants $k_1, k_2>0$,  
\[
\mathcal{H}\subset \mathcal{P}=\{p\in C(\R,\mathbb{C})\mid \|p\|_\infty \leq k_1,\; \rm{Lip}(p)\leq k_2 \}\,,
\]
for the standard sup-norm 
and  Lipschitz-norm, 
and $\mathcal{P}$ contains random bounded perturbations of the elements in $\mathcal{H}$ as well as functions which are not KBM. 
This construction permits to apply deterministic methods to analyse  the dynamics of ``almost stochastic equations", that is, equations with a time-dependent coefficient subject to a random variation. This is the case of the coefficient  ${\rm Im}(p_\Theta(t))$ in \eqref{eq:m2}, given by the imaginary part of a map in $\mathcal{H}_0$.

\section{Conclusions}
Our study investigates a nonautonomous, aperiodic energy balance model (EBM), employing the skew-product framework alongside recent advances in the theory of scalar nonautonomous concave and coercive differential equations. This approach enables a rigorous characterization of the model’s dynamics, revealing the existence of two attracting and one repelling hyperbolic solutions. Notably, despite its simplicity, the model captures qualitative features of global atmospheric temperature variability that closely mirror observed data. Crucially, the nonautonomous formulation introduces genuine dynamical complexity---permitting bifurcations that give rise to behaviors such as almost automorphic dynamics or chaos, which are excluded in the autonomous case but potentially relevant near critical climate transitions. 

A key outcome of our analysis is the identification of substantial limitations in approximating nonautonomous EBMs with autonomous models. We provide a rigorous application of the averaging method to highlight these pitfalls---often overlooked in the literature. In particular, the presence of chaotic and stochastic influences in climate systems undermines the theoretical justification for autonomous approximations.

Nevertheless, when averaging theory is applicable, we demonstrate how hyperbolicity can be leveraged to extend the validity of error estimates between the nonautonomous and averaged models. Specifically, within suitable neighborhoods of hyperbolic trajectories, these estimates can be prolonged from compact time intervals to the entire positive half-line. Furthermore, we show that hyperbolic solutions persist under small stochastic perturbations, allowing for a quantitatively controlled extension of the error estimates even when averaging theory no longer applies.

This work provides a comprehensive and rigorous account of the dynamics of zero-dimensional nonautonomous EBMs, including under climate forcings consistent with IPCC’s Socio-Economic Representative Pathways for the 21st century. The results underscore the necessity and potential of nonautonomous methods in theoretical climate science and aim to encourage further adoption of such approaches in the field.

\subsubsection*{Acknowledgments} We thank two anonymous referees for their precious feedback.  All authors were partly supported by MICIIN/FEDER project PID2021-125446NB-I00 and by the University of
Valladolid under project PIP-TCESC-2020. I.P. Longo was also partly supported by UKRI under the grant agreement
EP/X027651/1.


\end{document}